\documentclass[12pt]{article}
\usepackage{amsbsy}
\usepackage{amsfonts}
\usepackage{amsmath}
\usepackage{amssymb}
\usepackage{apacite}
\usepackage{setspace}
\usepackage{graphicx}
\usepackage{bm,multicol}
\usepackage[english]{babel}
\usepackage{natbib}
\usepackage[T1]{fontenc}
\usepackage[utf8]{inputenc}
\usepackage{authblk}
\usepackage{xcolor}

\newcommand{\newc}{\newcommand}
\newc{\N}{\mbox{N}}
\newc{\1}{\bf{1}}

\begin{document}
\title{A Latent Variable Approach for Modeling Relational Data with Multiple Receivers}

\author{J. Mulder, \& Peter D. Hoff}

\maketitle

\begin{abstract}
Directional relational event data, such as email data, often contain unicast messages (i.e., messages of one sender towards one receiver) and multicast messages (i.e., messages of one sender towards multiple receivers). The Enron email data that is the focus in this paper consists of 31\% multicast messages. Multicast messages contain important information about the roles of actors in the network, which is needed for better understanding social interaction dynamics. In this paper a multiplicative latent factor model is proposed to analyze such relational data. For a given message, all potential receiver actors are placed on a suitability scale, and the actors are included in the receiver set whose suitability score exceeds a threshold value. Unobserved heterogeneity in the social interaction behavior is captured using a multiplicative latent factor structure with latent variables for actors (which differ for actors as senders and receivers) and latent variables for individual messages. A Bayesian computational algorithm, which relies on Gibbs sampling, is proposed for model fitting. Model assessment is done using posterior predictive checks. Numerical simulations show that the model is widely applicable for various scenarios involving multicast messages. Furthermore, a mc-amen model with a 2 dimensional latent variable can accurately capture the empirical distribution of the cardinality of the receiver set and the composition of the receiver sets for commonly observed messages in the Enron email data. In the Enron network, actors have a comparable (but not identical) role as a sender and as a receiver in the network.
\end{abstract}


\section{Introduction}
Social behavior between individuals in a network can often be characterized by short communication messages of one actor towards another actor or several other actors. In an information network of employees in an organization a relational event can be an email message sent by an employee to one or several fellow employees (e.g., \cite{Perry:2013,MulderLeenders:2019}), in a classroom a relational event can be a teacher hushing one or several students (\cite{DuBois:2013}), and in a military convoy a relational event can be a soldier giving a command to one or more fellow soldiers (\cite{Leenders:2016}). Such relational data are becoming increasingly available due to the technical innovations of email, instant messaging apps, or radio communication, which have resulted in further development of statistical models for time-stamped social interaction data (e.g., \cite{Butts:2008,Brandes:2009,Quintane:2014,Stadtfeld:2017,meijerink2022discovering,meijerink2022dynamic,arena2022bayesian}).

\begin{figure}[t]
\begin{center}
\makebox{\includegraphics[width=8.0cm]{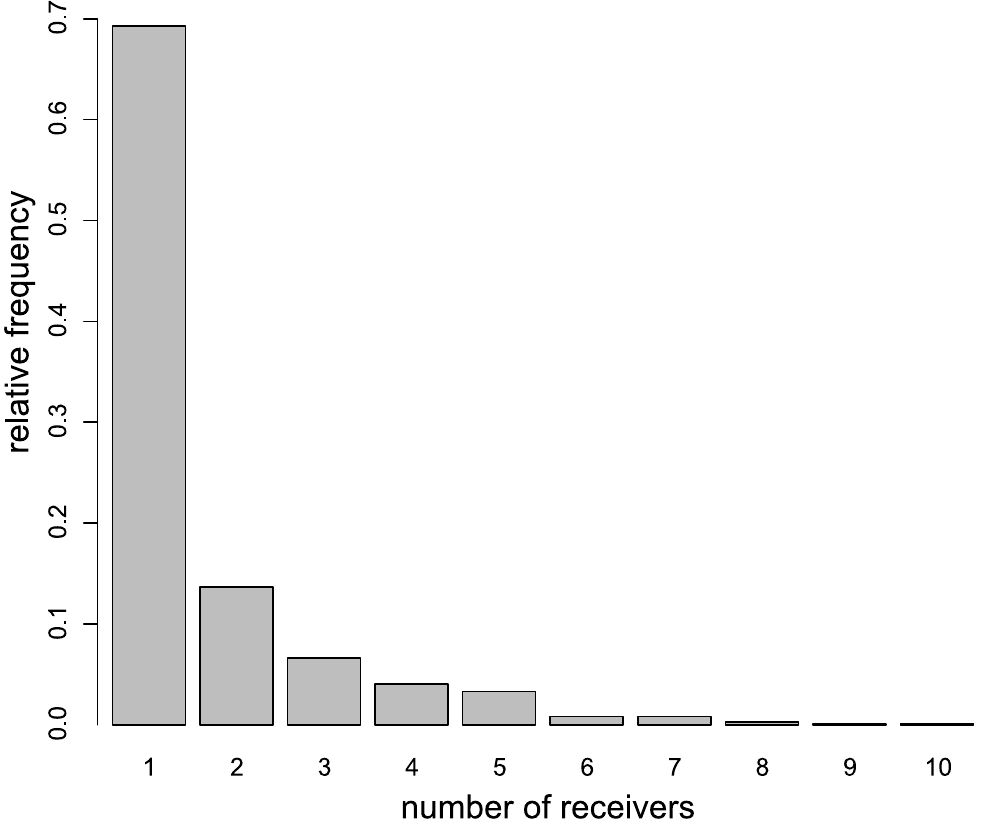}}
\end{center}
\caption{Empirical distribution of the number of receivers per message in the Enron email data \cite{cohen2009enron,zhou2007strategies}.}
\label{hist_obs_m}
\end{figure}

A key property of these relational data is that they contain unicast messages (i.e., messages sent from one actor to another actor) as well as multicast messages (i.e., messages sent from one actor to multiple other actors). The Enron email data (\cite{cohen2009enron}) as compiled by \cite{zhou2007strategies} with $M=21635$ messages among $N=156$ actors, which is the focus in this paper, consists of 69\% unicast messages and 31\% multicast messages\footnote{Data available at \url{https://github.com/jomulder/enronemaildata}.}. The empirical distribution of the cardinality of the receiver set over the entire data is presented in Figure \ref{hist_obs_m}. The distribution of the number of receivers has a typical shape that is often observed with email data in practice (e.g., see also the email data from \cite{eckmann2004entropy} and \cite{MulderLeenders:2019}) where unicast messages are observed most frequently, and the observed number of multicast messages tends to decrease as the size of the receiver set grows.


\begin{figure}[t]
\begin{center}
\makebox{\includegraphics[width=8.0cm]{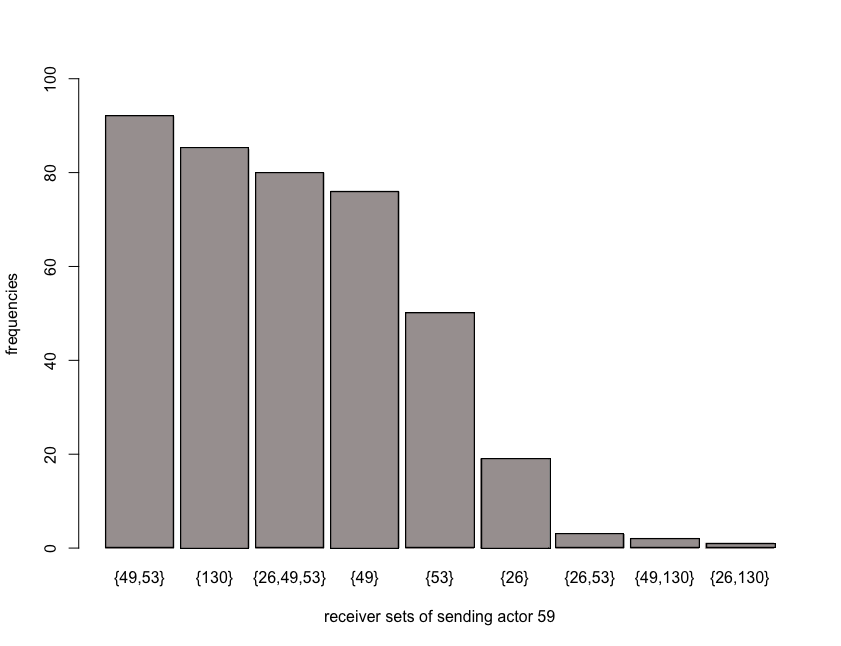}}
\end{center}
\caption{Observed frequencies of the receiver sets of messages sent by actor 59 in a subset of the Enron data consisting of actors 26, 49, 53, 59, and 130.}
\label{freq_sub59}
\end{figure}

Even though multicast relational data have not received a lot of attention in the statistical literature, multicast messages reveal important insights about social relations between actors, about social hierarchy of actors, and about the unique role of actors in a network. To make this more concrete, we consider a subset of the messages sent between actors 26, 49, 53, 59, and 130 in the Enron email network (all from the legal department) who sent 24, 77, 13, 408, and 102 messages among each other, respectively, out of which 70.5\% are unicast messages and 29.5\% multicast messages (similar as in the full data set). The observed frequencies of the receiver sets of the most active sender (actor 59), with 408 sent messages, can be found in Figure \ref{freq_sub59}. First we see that actor 130 is mainly popular as a receiver in unicast messages and this actor is practically absent in multicast messages. This suggests that actor 130 has a unique relation with this sender which is different from the other actors. It may be that actor 59 and actor 130 are both involved in a specific project or topic which is not relevant for the other actors. Second, we see that actors 49 and 53 have a similar role in terms of their relative popularity in unicast messages (where actor 49 is slightly more popular) as well as in multicast messages (where they generally appear both in frequently observed receiver sets, namely \{49,53\} and \{26,49,53\}). This may suggest that actors 49 and 53 have a similar role or hierarchical level in the network in relation to actor 59.  Third, we can see that actor 26 has a somewhat subordinate role (relative to actor 49 and actor 53) as actor 26 is mainly popular as a receiver when actors 49 and 53 are also in the receiver set but actor 26 is not a popular receiver in unicast messages.

Because most statistical approaches for modeling relational data are designed for dyadic observations involving two actors, one could split all multicast messages into multiple unicast messages of the sender to the different receivers. Effectively we would then be left with the frequencies of how often each actor appears as a receiver for every given sender and thereby loose the unique information about the specific roles of actors contained in multicast messages. For example, for the above mentioned subset of messages that were sent by actor 59, this separation step of the multicast messages would have resulted in observed inclusion proportions of 0.25, 0.61, 0.55, and 0.22 for actors 26, 49, 53, and 130, respectively. Based on these proportions we can only conclude that actors 26 and 130 are less popular receivers than actors 49 and 53. We no longer see the unique role of actor 130, the similar roles of actor 49 and 53, or the subordinate role of actor 130 relative to actors 49 and 53. Another issue of the separation step is that no information is available any more about the empirical distribution of the number of receivers in the data (such as in Figure 1), making it problematic (or practically impossible) to capture this property of the data when fitting a statistical model. This separation step would also unduly blow up the number of observations resulting in a possible overestimation of our statistical certainty. For example, the 408 messages sent by actor 59 would result in 666 unicast messages after separation. Finally note that directly modeling all possible compositions of the receiver set of multicast messages for a given sender in a network of $N$ actors can be challenging as the number of compositions is huge (namely $2^{N-1}-1$ when excluding the empty receiver set). For this reason, modeling email data (or relational data in general) with multicast messages is a challenging problem in applied social network research.

To address this problem, this paper presents a multicast additive and multiplicative effects network (mc-amen) model for directly modeling unicast and multicast relational observations. The latent factor structure builds upon the amen model (\cite{hoff2015dyadic}). In the mc-amen model, for a given message of a given sender, all potential receivers are placed on a latent suitability scale where a higher (lower) suitability score makes it more (less) likely that the actor will be included in the receiver set in a message. Next, a threshold value is drawn from a truncated normal distribution that is bounded by the largest suitability score. Only the actors whose suitability score exceed the threshold value are included in the receiver set of the message. As will be shown, a truncated normal distribution for the threshold value results in the typical distributional shape of the number of receivers as in Figure 1. Moreover, the mc-amen model uses a multidimensional latent variable for the actors as well as for the messages to capture unobserved heterogeneity in the data. A multiplicative approach is considered due to its ability to capture unobserved interaction behavior between the actors as implied by a latent distance model (\cite{Hoff:2002}) and as implied by a stochastic block model (\cite{NowickiSnijders:2001,pmlr-v31-dubois13a}), e.g., see \cite{Hoff:2008,Hoff:2009}. As the interaction behavior of actors be be different as a sender and as a receiver, each actor has a the latent variable as a receiver and a latent variable as sender. When viewing each dimension of a latent dimension as a different topic (or working project), these latent variables can be seen as a quantification of an actors' interest or involvement in this topic, either as a sender or as a receiver. Furthermore, the latent variables of the messages then quantify the degree of whether each topic is covered in a given message. Consequently an actor is likely to become a receiver for a given message if the latent variables of this potential receiver, the latent variables of the sender, and the latent variables of the message are all large (in absolute manner) and have an equal sign. Finally, the normal error is added to the latent suitable scores, similar as a probit regression model, to allow a straightforward posterior computation using Gibbs sampling in a Bayesian framework.

The mc-amen model avoids potential limitations of related approaches for multicast messages. \cite{Perry:2013} extended the Cox proportional hazard model to relational observations using a partial likelihood formulation for the actors in a receiver set for a given sender, the observed time, and the number of receivers in a message. When conditioning on the cardinality of the receiver set, it is implicitly assumed that an actor first chooses the size of the receiver set and then chooses who to include as receivers in the message. This does not seem to be a realistic mechanism in real life. Furthermore their approach does not provide a data generative mechanism for the number of receivers in a message, which is a key stochastic property in the data. Furthermore, their approach does not include any latent variables to capture potential unobserved heterogeneity. The mc-amen model on the other hand provides a data generative mechanism of both the size of the receiver set and the actors that are included in the receiver set. A flexible latent variable approach is implemented to capture unobserved heterogeneity regarding the interaction behavior between actors.

Furthermore, \cite{Shafiei:2010} proposed a mixed membership stochastic block model for the inclusion probabilities of the actors in the receiver set of a message sent by a given sender. In their latent variable approach, actors have the same latent variable as sender and receiver. This may not always be realistic however. For example, a manager may be likely to receive updates from his/her team members about a certain topic or working project just to stay informed but he/she may not be likely to send messages about this working project him/herself. Moreover, by not considering latent variables for the messages, the potential variability of topics in the messages is not captured. As will be shown in this paper, the latent variables of the messages are effective to capture the composition of the receiver sets in multicast messages. Another potential limitation of their approach is that the use of inclusion probabilities for each potential receiver results in a strictly positive probability of an empty receiver set. This contrasts with the observed data which only consists of receiver sets with at least one receiver. These limitations are automatically handled by the mc-amen model.

The paper is organized as follows. Section 2 presents the multiplicative latent factor model for relational data with multiple receivers. To fit the model a Bayesian implementation is considered. Prior specification and posterior predictive checks for model assessment are discussed. Next, Section 3 describes numerical simulations to assess the behavior of the mc-amen model under different scenarios. Section 4 presents two analyses of the Enron email network. First, an empirical analysis is discussed of the subset of actors 26, 49, 53, 59, and 130 that was also discussed in this introduction. Second, an empirical analysis of the full data set is discussed. The goal of the  empirical analyses is to assess how well the mc-amen model can capture (i) the empirical distribution of the cardinality of receiver sets of the entire data set (Figure \ref{hist_obs_m}) and (ii) the multicast interaction behavior between actors in the Enron network. We end the paper with a discussion in Section 5.

\section{A multiplicative latent factor model for relational events with multiple receivers}
We start with some notation. A network is considered for a set of actors denoted by $\mathcal{A}$ of size $|\mathcal{A}|=N$. 
The binary vector $\textbf{y}_{ai}$ of length $N$ captures the receiver set of the $i$-th message sent by actor $a$ actor where $y_{aia'}=1$ if actor $a'$ is in the receiver set, and $y_{aia'}=0$ if $a'$ is not in the receiver set, for $i=1,\ldots,m_a$, where $m_s$ is the number of messages sent by $s$. The total number of messages in the data is denoted by $M=\sum_{a=1}^N m_a$. The receiver set of message $i$ of actor $a$ is then written as $\mathcal{R}_{ai}=\{a'|y_{aia'}=1\}$. We assume that an actor cannot send a message to one's self even though the model would allow for this if needed. Also note that messages sent to one's self do not contain any information about the relationships between actors.

We propose the following generative model for the binary receiver vector $\textbf{y}_{ai}$ for message $i$ sent by actor $a$,
\begin{eqnarray}
\nonumber \textbf{z}_{ai,-a} &\sim &N(\bm\theta_{ai,-a},\textbf{I}_{N-1}),\\
\label{model1} c_{ai} | \textbf{z}_{ai,-a} &\sim& TN(\mu_c,\sigma^2_c,-\infty,z_{ai(N)}),\\
\nonumber y_{aia'}&=&\left\{\begin{array}{ll}
1 & \text{if }z_{aia'}>c_{ai}\\
0 & \text{elsewhere,}\end{array}\right.
\label{model1a}
\end{eqnarray}
where the latent variable $z_{aia'}$ quantifies the suitability of message $i$ for potential receiver actor $a'$ sent by actor $a$ on a latent suitability scale, $\theta_{aia'}$ is the mean suitability, $c_{ai}$ is a threshold parameter for message $i$ sent by $a$ which follows a truncated normal distribution with mean $\mu_c$ and variance $\sigma^2_c$ in the subspace $(-\infty,z_{ai(N)})$, and where $z_{ai(N)}$ is the $N$-th ordered value of the latent suitability scores, i.e., $z_{ai(N)}=\max_{a'} \{z_{aia'}\}$. By using a truncated distribution with upper bound $z_{ai(N)}$ the receiver set will never be empty. The mean $\mu_c$ and the variance $\sigma_c^2$ of the distribution of the threshold parameters control the distribution of the number of receivers across messages.

An additive and multiplicative latent effects model is considered for the mean suitability score according to
\begin{eqnarray}
\theta_{aia'} &=& \textbf{x}_{aia'}^{\top}\bm\beta + b_{a'} + \textbf{u}_{a'}^{\top}\textbf{v}_a + \textbf{u}_{a'}^{\top}\textbf{w}_{ai},\\
b_{a'} &\sim & N(\mu_b,\sigma^2_b)
\label{model1b}
\end{eqnarray}
where $\textbf{x}_{aia'}$ is a vector of $K$ observed predictor variables (e.g., exogenous or endogenous), $\bm\beta$ contains the $K$ unknown coefficients that quantify the relative importance of these predictor variables, the random effect $b_{a'}$ quantifies the average popularity of actor $a'$ as a receiver, which follows a normal distribution with mean $\mu_b$ and variance $\sigma^2_b$, $\textbf{u}_{a'}$ is a vector containing the $Q$ latent variables of actor $a'$ as a receiver, $\textbf{v}_a$ is a vector containing the $Q$ latent variables of actor $a$ as sender, and $\textbf{w}_{ai}$ is a vector containing the $Q$ latent variables of message $i$ by sender $a$. 

Each latent variable can be viewed as a theme or topic on which an actor can be (in)active as a sender or as a receiver, or which is (not) important in a message. If a latent variable of actor $a$ as sender, $v_{aq}$, is large in absolute value and of the same (opposite) sign as the latent variable of a potential receiver actor $a'$, $u_{a'q}$, it becomes likely (unlikely) for the sender to include this potential receiver in the receiver set. Furthermore the inner product of the latent vectors imply that the multiplicative function is added for the different latent dimensions, i.e., $\textbf{u}_{a'}^{\top}\textbf{v}_a=u_{a'1}v_{a1}+\ldots+u_{a'Q}v_{aQ}$. Similarly, if a latent variable of message $i$, $w_{aiq}$, is large in absolute sense and of the same (opposite) sign as the latent variable of a potential receiver actor, $u_{a'q}$, it becomes likely (unlikely) to include this potential receiver in the receiver set of message $i$. By also including a latent variable for each message, the suitability scores of the potential receivers are not conditionally independent given the latent variables of the sender and the potential receivers, the fixed effects, and the random popularity effects. 
Finally note that by assuming different latent variables of each actor as a receiver and as a sender, the model allows for different roles of an actor as a receiver and as a sender on the latent topics. Based on the fitted model to the data at hand, it can be investigated of course how strong the latent variables of the actors as senders and as receivers correlate with each other. 

\begin{figure}[t]
\centering
\makebox{\includegraphics[width=10.0cm]{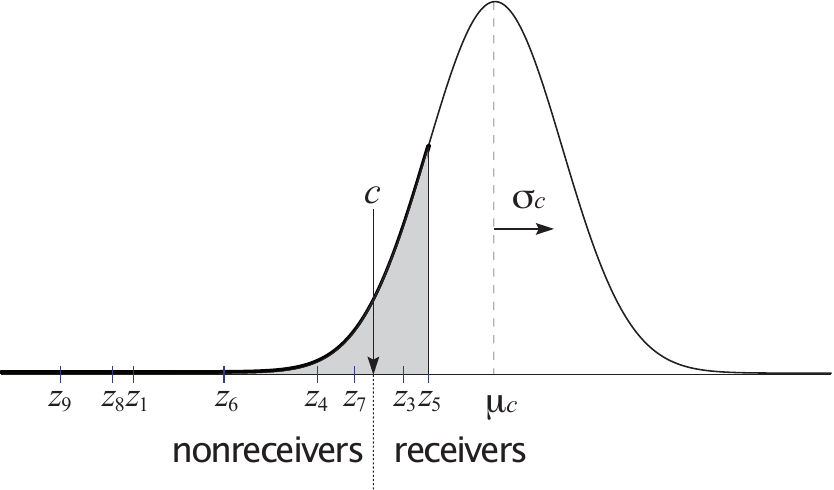}}
\caption{Graphical representation of the inclusion of the actors 3 and 5 in the receiver set for a message send by actor 2 based on the random threshold value $c$. The index of the sender actor $a$ and message $i$ are omitted.}
\label{figinclusion}
\end{figure}

In sum, the model builds on the following intuition. When an actor decides to send a message, the sender actor determines the mean suitability for all potential receivers for the given message which is based on observed predictor variables (e.g., observed past interactions between actors or common locations where the actors work), the average popularity of the actors as a receiver, the multiplicative function of the latent variables of the sender and the potential receiver, and the multiplicative function of the latent variables of the message and the potential receiver. Random normal errors with a variance of 1 are added to the suitability scores. The suitability scores are placed on a latent scale. Subsequently, a threshold value is drawn from a truncated normal distribution where the upper bound equal to the largest suitability score. This ensures that the receiver set will always contain at least one actor, and the actor with the largest suitability score will always be included (Figure \ref{figinclusion}). All actors are included in the receiver set whose suitability score exceeds the drawn threshold value. Given the suitability scores of potential receivers, $\textbf{z}_{ai,-a}$, the inclusion probability for actor $a'$ in the receiver set $\mathcal{R}_{si}$ is then given by
\[
\text{Pr}(r\in R_{ai}|\textbf{z}_{ai,-a},\mu_c,\sigma_c) = \frac{\Phi(\frac{z_{aia'}-\mu_c}{\sigma_c})}
{\Phi(\frac{z_{ai(N)}-\mu_c}{\sigma_c})},
\]
where $\Phi(\cdot)$ denotes the standard normal cdf.

Because the mean of the threshold values, $\mu_c$, and the mean of the popularity parameters, $\mu_b$, both define the location of the latent suitability scale, we fix $\mu_c=0$ and allow $\mu_b$ to be freely estimated to avoid identifiability issues. Moreover, the variance of the threshold distribution, $\sigma_c^2$, is difficult to identify when the means of the popularity parameters, $\mu_b$, are freely estimated. This can be explained by the truncated normal distribution of the threshold parameter (having an increasing shape as in Figure \ref{figinclusion}) which remains virtually identical for specific combinations of the mean of the suitability scores and the standard deviation of the threshold distribution. To avoid any identifiability issues as a result of this, the standard deviation of the threshold distribution is fixed at $\sigma_c^2=1$.

Given this truncated normal distribution of the threshold parameter having a mean parameter of 0 and a variance of 1, the distribution of the cardinality of the receiver set will follow the typical decreasing shape we often see in email data with multicast messages (e.g., see Figure \ref{hist_obs_m} for the Enron email data) when the means of the suitability scores are all negative (as in Figure \ref{figinclusion}). Consequently, the popularity parameters, $b_a$, which serve as intercepts for the mean of the suitability scores of each actor, will be negative when fitting the mc-amen model to such data.


\subsection{Prior specification and posterior computation}\label{section_prior}
For all free parameters, vague conjugate priors are considered to facilitate easy posterior sampling and to ensure that the posterior is completely dominated by the data:
\begin{eqnarray*}
\bm\beta & \sim & \mathcal{N}(\bm\beta_0,\bm\Psi_0)\\
\sigma_b^2 & \sim & IG(\alpha_b,\gamma_b)\\
\sigma_c^2 & \sim & IG(\alpha_c,\gamma_c)\\
\textbf{v}_a, \textbf{u}_{a'}, \textbf{w}_{ai} &\sim & N(\textbf{0},\textbf{D}),
\end{eqnarray*}
where the hyperparameters are set to $\bm\beta_0=\textbf{0}$, $\bm\Psi_0=1\text{e}3\textbf{I}_K$, $\alpha_b=\gamma_b=\alpha_c=\gamma_c=1$\footnote{Based on our analyses, all results were virtually independent of the exact choice of the hyperparameters of the inverse gamma priors.}, and $\textbf{D}=D\textbf{I}_Q$. For the analysis of the Enron email data, a large prior variance is set for the latent variables of $D=1\text{e}3$.

The conditional posterior distributions for $\bm\beta$, $\textbf{b}$, $\textbf{u}_r$, $\textbf{v}_s$, and $\textbf{w}_{si}$ then follow multivariate normal distributions, the random effects variance $\sigma^2_b$ follows inverse gamma distribution, and the threshold parameters $c_{si}$ follow truncated normal distributions $N(\mu_c,\sigma_c^2)$ in the interval $(\max_{r:y_{sir}=0} \{ z_{sir} \},$ $\min_{r:y_{sir}=1} \{ z_{sir} \})$. Finally, the conditional posterior for the latent suitability scores is proportional to
\begin{align*}
\pi(\textbf{z}_{si,-r}|\bm\theta_{si}, \textbf{y}_{si,-r}, c_{si}) & \propto  N(\bm\theta_{si,-r},\textbf{I}_{P-1})
\times\\
&\prod_{r\not =s} 1(z_{sir}<c_{si})^{1-y_{sir}}1(z_{sir}>c_{si})^{y_{sir}}
\Phi\left(\frac{z_{si(P)}-\mu_c}{\sigma_c}\right)^{-1}.
\end{align*}
To sample $\textbf{z}_{si,-r}$, first a candidate $\textbf{z}_{si,-r}^*$ is drawn from a multivariate truncated normal $N(\bm\theta_{si,-r},\textbf{I}_{P-1})\prod_{r\not =s} 1(z_{sir}<c_{si})^{1-y_{sir}}1(z_{sir}>c_{si})^{y_{sir}}$, and the draw is accepted with probability $\min\left(\frac{\Phi((z_{si(P)}^{(l-1)}-\mu_c)/\sigma_c)}{\Phi((z_{si(P)}^*-\mu_c)/\sigma_c)},1\right)$, where $\textbf{z}_{si}^{(l-1)}$ is the previous (`$l-1$'-th) draw, using a Metropolis-Hastings step. These conditional distributions can be used to obtain a MCMC algorithm from which it is relatively straightforward to sample from, yielding posterior draws from the fitted model.

\subsection{Assessing model fit using posterior predictive checks}
Posterior predictive checks are flexible to assess whether a model captures certain key characteristics of the data (e.g., \cite{Meng:1994,Gelman,Kollenburg:2015}). Moreover in the presence of latent variables, posterior predictive checks are generally easier to apply (and perhaps easier to interpret) than traditional model fit indices which assess global fit, possibly by correctly for model parsinomy, such as the BIC. For this reason posterior predictive checks are used for evaluating the fit of the mc-amen model, and specifically to determine the number of dimensions of the latent variable for the actors and messages. Replicated data sets are generated from the posterior predictive distribution using the posterior draws of the unknown parameters from the fitted mc-amen model. These replicated data sets have the same sample size as the observed data. Next, it is assessed whether certain characteristics of interest, referred to as test statistics, are similar between the replicated data and the observed data. If the differences are deemed to be too large, this suggests that the model badly captures this characteristic of the observed data.

The following test statistics are used for the posterior predictive checks of the mc-amen model:
\begin{itemize}
\item The frequency of a specific combination of a sender and the receiver set in the observed data (either for unicast messages or multicast messages). This will be done for the most frequently observed combinations of senders and receiver sets.
\item The distribution of the cardinality of the receiver sets over the entire data set. This will show whether the distributional form of the size of the receiver sets in the observed data is well-captured by the model, such as the typical decreasing trend as in the Enron data in Figure \ref{hist_obs_m}.
\item The transitivity in the entire data set, given by
\begin{align*}
&t_1(\textbf{Y}) = \sum_{s\not = r}\sum_{j\in\mathcal{A}\backslash \{s,r\}} {e}_{s,j}{e}_{j,r}{e}_{s,r},\\
&t_2(\textbf{Y}) = \sum_{s\not = r}\sum_{j\in\mathcal{A}\backslash \{s,r\}} {e}_{s,j}{e}_{j,r}{e}_{r,s},
\end{align*}
where $e_{s,r} = \sum_{i} y_{s,i,r} - \frac{1}{P-1}\sum_{i,r} y_{s,i,r}$, i.e., the difference between the total number of messages sent by $s$ to $r$ receiver and the average number of receivers of messages sent by $s$. For the current paper we consider a posterior predictive check for transitivity which captures the tendency of actor $s$ to send messages to actor $r$ as a function of the number of messages sent by $s$ to other actors than $r$ and the number of messages sent by these other actors to $r$. Transitivity is often phrased as ``the friends of my friends are my friends'', an example of structural balance theory (\cite{Heider:1946,Cartwright:1956}). Transitivity can also be seen as a form of ``broker-skipping'' as it captures the  tendency of actors to skip intermediate actors in a communication network (e.g., \cite{Leenders:2016}). Multiplicative latent factor models, such as the mc-amen model, are known to capture such higher-order dependency structures in the data (\cite{Hoff:2005,Hoff:2009}).
\end{itemize}
Note that the first quantity, i.e., the number of observed messages of a specific sender to a specific \textit{receiver set}, is preferred over the number of messages of a specific sender to a specific receiver based on all receiver set for this given sender, i.e., the relative popularity of an actor as a receiver for a given sender, when assessing model fit for relation with with multicast messages. For example, if actor 1 had send 100 unicast messages to actor 2, 100 unicast messages to actor 3, and no multicast messages to actors 2 and 3, the relative popularity of actors 2 and 3 would be the same for this sender as when actor 1 had send 100 multicast messages to both actors 2 and 3, and no unicast messages to actors 2 and 3 separately. Thus, by solely looking at the relative popularity, we would not be able to see whether the multicast behavior of actors in the observed data is well captured by the model.

\section{Numerical simulations}
\subsection{Simulation 1. Parameter recovery and model fit under different populations}\label{simulations1}
Numerical simulations were performed to get insights about the recovery of the parameters and the performance of the model fit indicators. A network of 30 actors was considered having a 1-dimensional latent variable. Table \ref{tabExtraSim} (Population 0, first row) shows the parameter values in a reference population, and the parameter values for certain alternative populations. In Population 1a and 1b, the variances of the latent variable of the sender and the messages were different (either var$(v_a)\ll $ var$(w_{a,i})$ or var$(v_a)\gg $ var$(w_{a,i})$). In Population 2a and 2b, the correlations of the latent variables of actors as a sender and receiver differed (either cor$(u_a,v_a)=.7$ or cor$(u_a,v_a)=-.7$). In Population 3, the effects of the covariates was increased with a factor of 4 making the latent variables relatively less important. In Population 4, there were 10 out of 30 actors having a larger standard deviation for the threshold parameter (namely $\sigma_c=5$) resulting in considerably larger receiver sets for these 10 sending actors. Thus, under Population 4, the model is misspecified as the variance of the threshold parameters is fixed at 1 for all actors under the mc-amen model. From each population, 25, 50, or 150 observations were generated from every sending actor. The covariates were generated from standard normal distributions. In the generated data, the proportion of multicast messages varied between 55\% and 75\% across the different populations. Models were fitted with $Q=0,$ 1, or a 2 dimensional latent variable. To assess model fit, the the posterior predictive distributions were considered for the frequencies of the 20 most commonly observed combinations of senders and receiver sets were considered as well as the frequencies of the cardinality of the observed receiver sets. When fitting the models, we did not observe any label switching for the latent variables with for the model with a 2 dimensional latent variable. Note that the labels of the two dimensions are arbitrary for this model type.

The 95\% credibility intervals of the parameters and latent variables and posterior predictive distributions can be found in Appendix A. The model with $0$, 1, and 2 latent variables are displayed in black, red, and green, respectively. For Population 0 to Population 4, Figures \ref{CrI_an1}-\ref{CrI_an7} shows the credibility intervals of the fixed effects, the random popularity parameters, and the latent variables of the actors as senders and receivers. Figures \ref{vu_an1}-\ref{vu_an7} shows the posterior means of the latent variables as a sender and as a receiver for the first latent dimension and for the second latent dimension (which is only available if for the model with a 2 dimensional latent variable). Finally, the posterior predictive distribution of the cardinality of the receiver sets and the frequencies of the 20 most commonly observed messages can be found in Figures \ref{cardi_an1}-\ref{cardi_an7} and Figures \ref{recset_an1}-\ref{recset_an7}, respectively. Under the Population 0, we can see that the parameters can be well-recovered (Figure \ref{CrI_an1}). Furthermore, we see that all three models (with $Q=0$, 1, and 2 dimensional latent variables) are able to capture the distribution of cardinality of the receiver set in the generated data (Figure \ref{cardi_an1}). Furthermore, Figure \ref{recset_an1} shows that the (misspecified) model with no latent variables cannot capture the observed frequencies of the 20 most commonly observed messages while the (correct) model with a 1 dimensional latent variable results in a satisfactory fit, as well as the model with a 2 dimensional latent variable shows hardly any improvement. Based on these plots, we would conclude that the model with 1 a dimensional latent variable is preferred (as the fit is not improved for the mode with a 2 dimensional latent variable).

Under Population 1a and 1b, the results are very similar to the results under Population 0. This implies that different variances of the latent variables can be well-captured by the model (Figures \ref{vu_an2} and \ref{vu_an3}). Under Population 2a and 2b, the results are also comparable. Thus, positive and negative correlations between the latent variables can be well captured by the model. Under Population 3, all three models (with $Q=0$, 1, or 2) fit the data approximately equally well (Figure \ref{recset_an2}). This can be explained by the fact that the latent variables have a much smaller impact in comparison to the fixed effects. In this situation, the model with 0 latent variables would be preferred. Finally under Population 4, the messages sent by the 10 actors with $\sigma_c=5$ had receiver sets with an average size of 11 and a standard deviation of 10, and the messages sent by the other 20 actors with $\sigma_c=1$ had receiver sets with an average size of 1.4 with a standard deviation of .8. Figure \ref{recset_an7} shows that the model with a 0 or 1 dimensional latent variable results in a poor which can be explained by model misspecification. Interestingly however the model with an additional latent dimension ($Q=2$) results in an acceptable fit. This implies that the additional latent dimension is able to pick up on this specific characteristic of the data. This can also be seen from Figure \ref{vu_an7} which shows that the estimated latent variables of the actors as sender ($v$) are clustered for these 10 actors and for the other 20 actors for one dimension under the model with $Q=2$. Furthermore, the model with a 2 dimensional latent variable is also able to capture the distribution of the cardinality of the receiver sets unlike the model with 0 or a 1 dimensional latent variable (Figure \ref{cardi_an7}). In conclusion, these simulations show that the mc-amen model is applicable to a wide range of populations.

\begin{table}[t]
\begin{center}
\caption{Population values of the parameters for a network of 30 actors with a 1-dimensional latent variable. For empty cells, the parameters are identical to the reference population (Population 0, first row). In Population 4, the standard deviation of the threshold parameter was 1 for 20 actors and 5 for 10 actors.}
\hspace*{-1.5cm}
{\small 
\begin{tabular}{lccccccccccc}
  \hline
                        & $\bm\beta$ & $(\mu_b,\sigma_b)$ & var$(u_{1,a})$ & var$(v_{1,a})$ & cor$(u_{1,a},v_{1,a})$ & var($w_{1,a,i})$
                        & $\sigma_c$\\
Population 0 & $(-\tfrac{1}{2},-\tfrac{1}{4}0,\tfrac{1}{4},\tfrac{1}{2})$ & (-7,.25) & 1 & 1 & 0 & 1 & 1\\
Population 1a & & &  & .1 &  & 10 & \\
Population 1b & & &  & 10 &  & .1 & \\
Population 2a & & &  &  & .7 &  & \\
Population 2b & & &  &  & $-.7$ &  & \\
Population 3 & $(-2,-1,0,1,2)$ & &  &  &  &  & \\
Population 4 & & &  &  &  &  & 1 or 5 \\
\hline
\end{tabular}}
\label{tabExtraSim}
\end{center}
\end{table}

\subsection{Simulation 2. Simulations based on the Enron email network}
A simulation study was conducted to show the performance of the mc-amen model for analyzing network data with multicast messages and to determine the dimensionality of the latent variable using posterior predictive checks under a controlled situation. Network data were generated based on the characteristics of the Enron email data which consisted of 156 actors who sent a total of 21,635 messages among each other varying from 0 to 1,597 with a median of 56 messages. To assess the effect of the sample size, three different sample sizes were considered in the simulation: the same number of messages as sent by the actors in the Enron data, half the number of messages as sent by the actors in the Enron data, and a quarter of the number of messages as sent by the actors in the Enron data. A 2-dimensional latent variable was considered for the actors and messages where the latent variables of the actors as a receiver were generated using a bivariate normal distribution with zero means and a variances of 1.8 and a covariance of 0.9, the latent messages of the actors as sender were independently generated from a uniform distribution in the interval $(-1,1)$, and the latent message variables were independently drawn from a discrete distribution having a value of 0 or 0.5 both with a probability of .5. These different distributions were considered to check how well the model can handle such different types of latent variables. Finally, we consider a mc-amen model with 32 coefficients (fixed effects) varying from -0.7 to 0.5 for $\beta_1$ to $\beta_{32}$, respectively, which is the same as the mc-amen model we consider for the Enron email data. Finally the random popularity parameters were generated using a normal distribution with mean $-8.7$ and a standard deviation of 1. This resulted in a distribution of the cardinality of the receiver sets in the generated data that is similar as in the Enron data (Figure 1). Thereby the simulation serves as a proof of concept for the analysis of the entire Enron email data. 

\begin{figure}[t]
\centering
\makebox{\includegraphics[width=12.0cm]{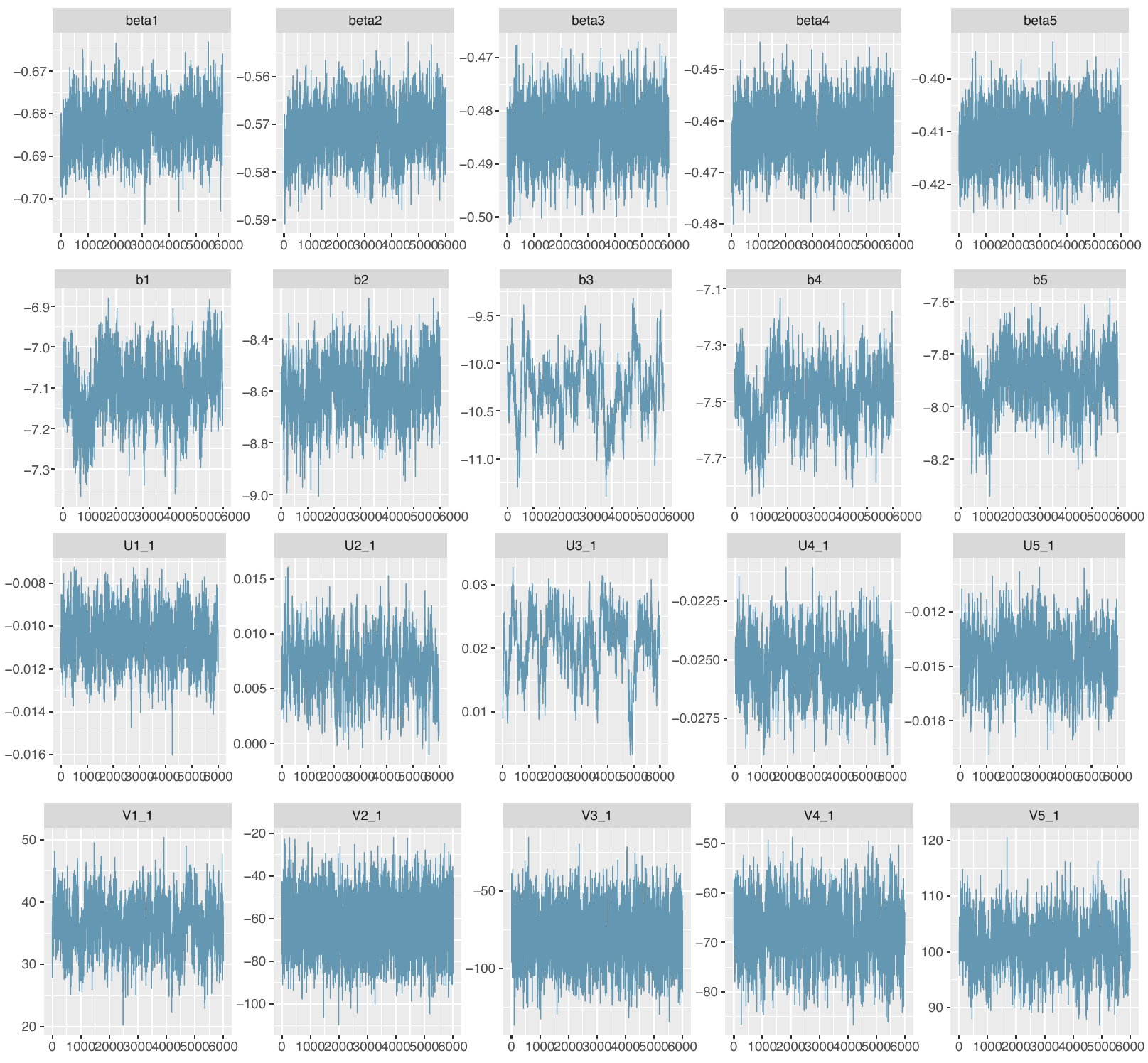}}
\caption{Trace plots of the posterior draws for a model with a latent variable with $Q=0$, 1, 2, and 3 dimensions fitted to the simulated data.}
\label{fig_sim_trace}
\end{figure}

For all three generated data sets, 600,000 posterior draws were obtained after burnin of which every 100th draw was stored to correct for autocorrelations. For one generated sample, Figure \ref{fig_sim_trace} shows the posterior draws of the first five $\beta$ coefficients, and the five popularity parameters and latent variables as sender and receiver of the first five actors based on a fitted model with a 3-dimensional latent variable for one data set \cite[plots were made using the `bayesplot' package][]{gabry2017bayesplot}. Overall the trace plots show convergence. The trace plots for the other parameters and other data sets looked similar.

\begin{figure}[t]
\centering
\makebox{\includegraphics[width=10.0cm]{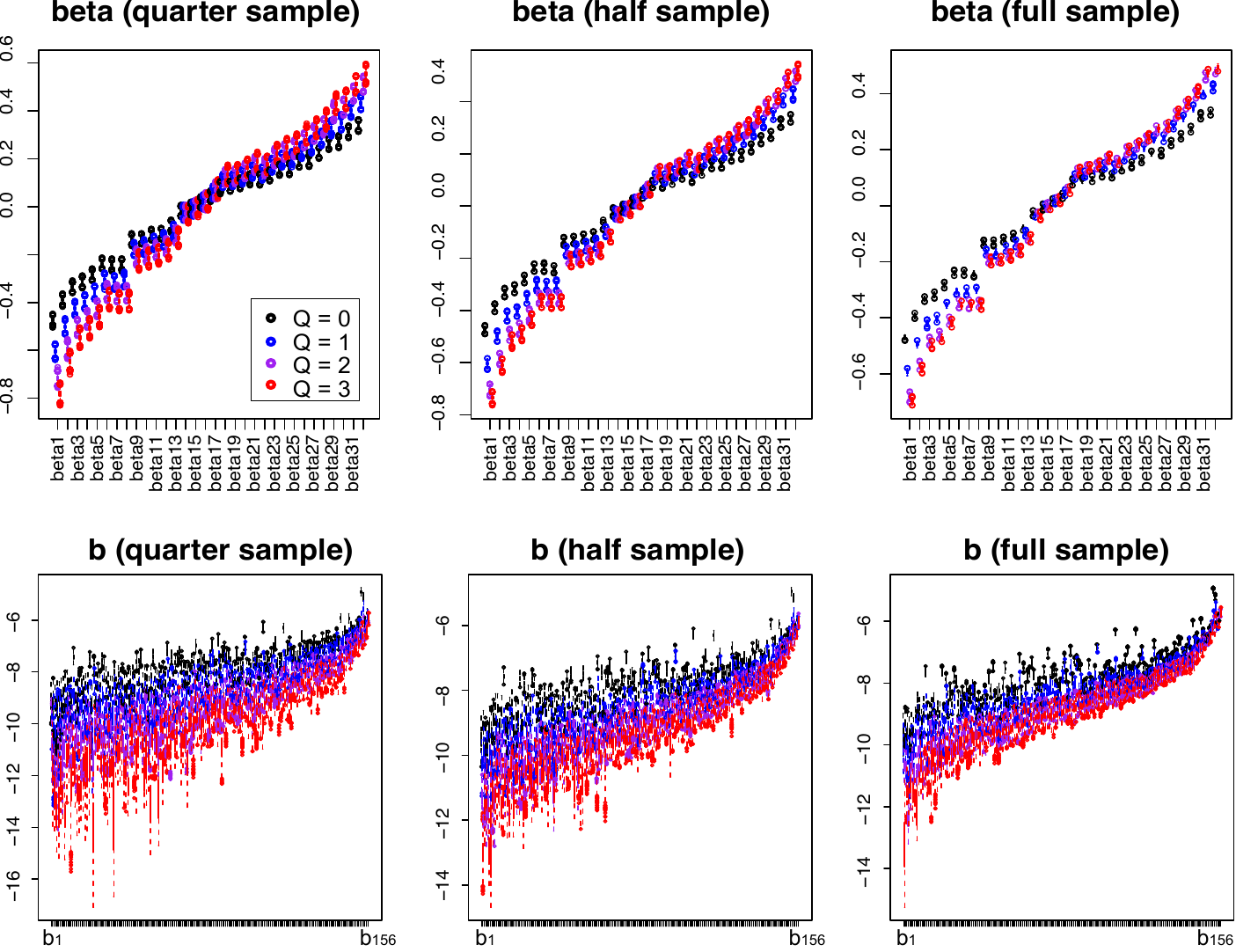}}
\caption{Posterior credibility intervals for the coefficients $\beta_1,\ldots,\beta_{32}$ (upper panels) and the popularity parameters $b_1,\ldots,b_{156}$ (lower panels) with a latent variable with $Q=0$, 1, 2, and 3 dimensions fitted to the simulated data for a quarter of the sample size (left panels), half the sample size (middle panels), and the full sample size (right panels).}
\label{fig_sim_beta_bvec}
\end{figure}

Figure \ref{fig_sim_beta_bvec} shows the Bayesian credibility intervals of the 32 coefficients (top panels) and the 156 popularity parameters (lower panels, ordered by size) for a quarter of the sample size (left panels), half the sample size (middle panels), and the full sample size (right panels). For every parameter the intervals are displayed in batches of the fitted model with $Q=0$, 1, 2, and 3 dimensions for the latent variable, displayed in black, blue, purple, and red, respectively. Overall the plots show that the relative magnitude of the parameters is fairly consistent for the different models with $Q=0$, 1, 2, and 3 dimensions of the latent variable. We see that for larger dimensions the popularity parameter becomes smaller on average. This can be a consequence of the variance of the suitability scores which increases for larger dimensions of the latent variable (because the mean suitability is modeled using model variance components). By accordingly decreasing the means of the suitability scores (which is controlled the popularity parameters), a similar distributional for the cardinality of the receiver sets can be obtained. 

\begin{figure}[t]
\centering
\makebox{\includegraphics[width=13cm]{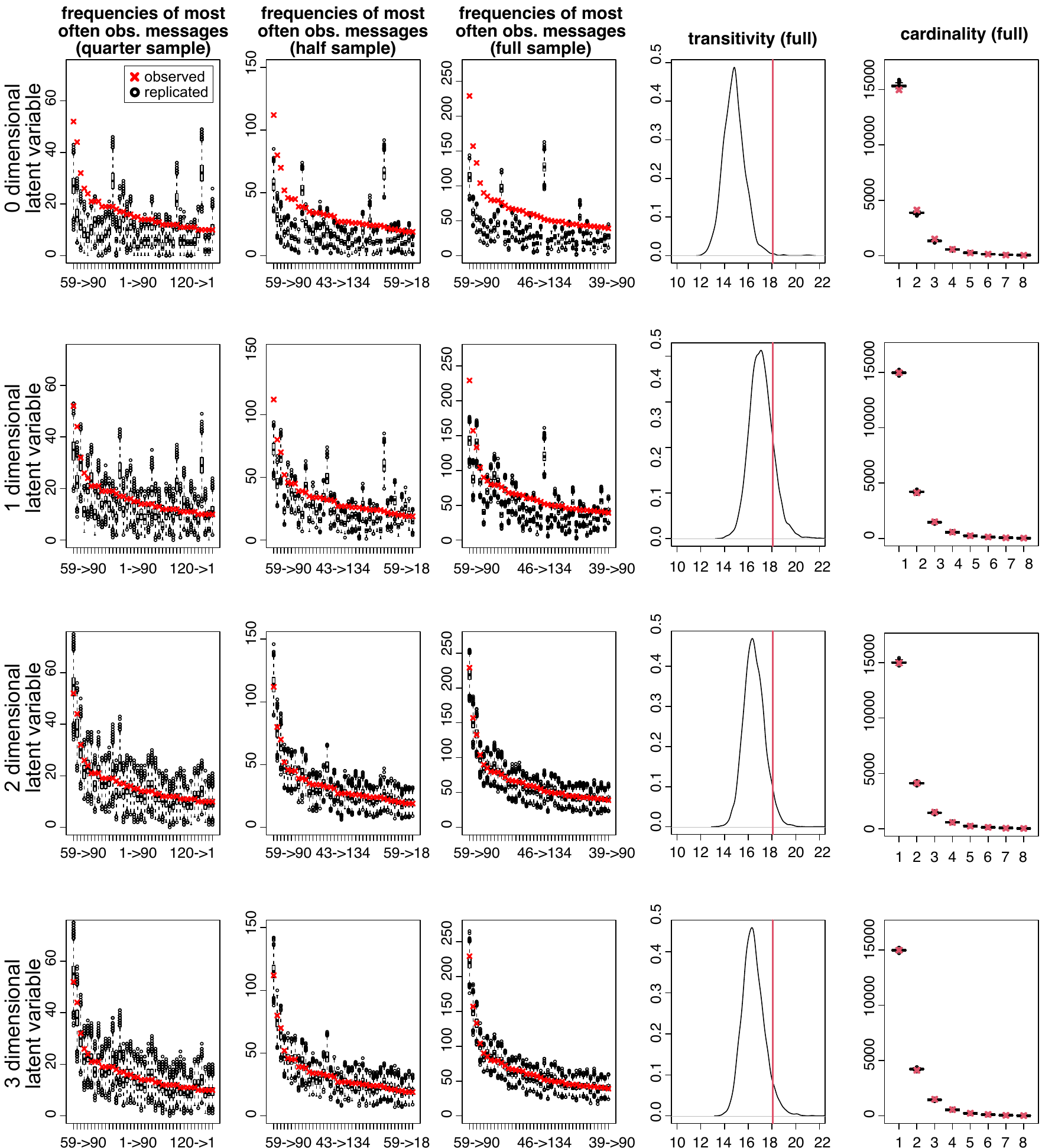}}
\caption{Posterior predictive distributions based on a model with a latent variable with $Q=0$, 1, 2, and 3 dimensions fitted to the simulated data. The red crosses and the vertical lines correspond to the respective quantities in the observed data.}
\label{fig_sim_pred}
\end{figure}

To assess model fit, predictive distributions of different statistical quantities were sampled under all fitted four models for all three data sets. To keep the presentation of the results as concise as possible, Figure \ref{fig_sim_pred} shows the posterior predictive distributions for the number of messages sent among the 40 most frequently observed combinations of senders and receivers for the three different sample sizes (quarter sample size in the panels in the first column, half the sample size in the second column, and the full sample size in the third column), the first transitivity for the full sample size (fourth column), and the distribution of the cardinality of the receiver set in the full data (fifth column). The plots of the transitivity and the cardinality for half the sample size and the quarter of the sample size looked similar, and are therefore omitted here. We see that the model without any latent variables (Figure \ref{fig_sim_pred}, first row) does not capture the observed frequencies very well. The first transitivity statistic is also poorly captured by this model (fourth column). The distribution of the cardinality of the receiver sets however is well captured by this model (fifth column) which illustrates that the mc-amen model is able to capture this property of the data very well even when the model shows a poor fit based on the frequencies. We see an improvement of the fit of the mc-amen with a 1-dimensional latent variable (second row) both in terms of the frequencies (in particular the first few most frequently observed dyads) and in terms of transitivity. There is still some misfit though based on the observed frequencies for certain messages. This misfit seems no longer present when a mc-amen model with a 2-dimensional latent variable is used (third row). This confirms our expectation because the data was also generated using a mc-amen with a 2-dimensional latent variable. Model fit only improves very slightly more when further increasing the latent variable to $Q=3$ dimensions (fourth row). Thus, to keep the model as parsimonious as possible, the mc-amen model with a 2-dimensional latent variable seem to be preferred for all three data sets.

\section{Analyses of the Enron email data}

\subsection{Analysis of a subset}
First we consider the subset of actors 26, 49, 53, 59, and 130 from the Enron email network discussed in the introduction in this paper. The goal of this analysis is to see whether the mc-amen model is able to properly capture the multicast behavior between the actors, to give some insights about the importance of latent variables to model multicast messages, and to show a comparison with existing approaches in a simple setup. No covariates were included in the analysis. In total 624 were send among these actors with most messages (408) sent by actor 59 (see the receiver sets and in Figure 2). We fit a mc-amen model with no latent variables, a mc-amen model with a 1-dimensional latent variable only for the actors and not for the messages (so without using $\textbf{w}_{ai}$), a mc-amen model with a 1-dimensional latent variable for all actors as well as messages, and a mc-amen model with a 2-dimensional latent variable for all actors and messages. The first model with no latent variables is comparable with the model of \cite{Perry:2013} which also does not contain any latent variables. The second model with a 1-dimensional latent variable for only the actors and not the messages is comparable with the model of \cite{Shafiei:2010} (their model was more restrictive however because the latent variables of the actors were equal as sender and as receiver). 

\begin{figure}[t]
\centering
\makebox{\includegraphics[width=12cm]{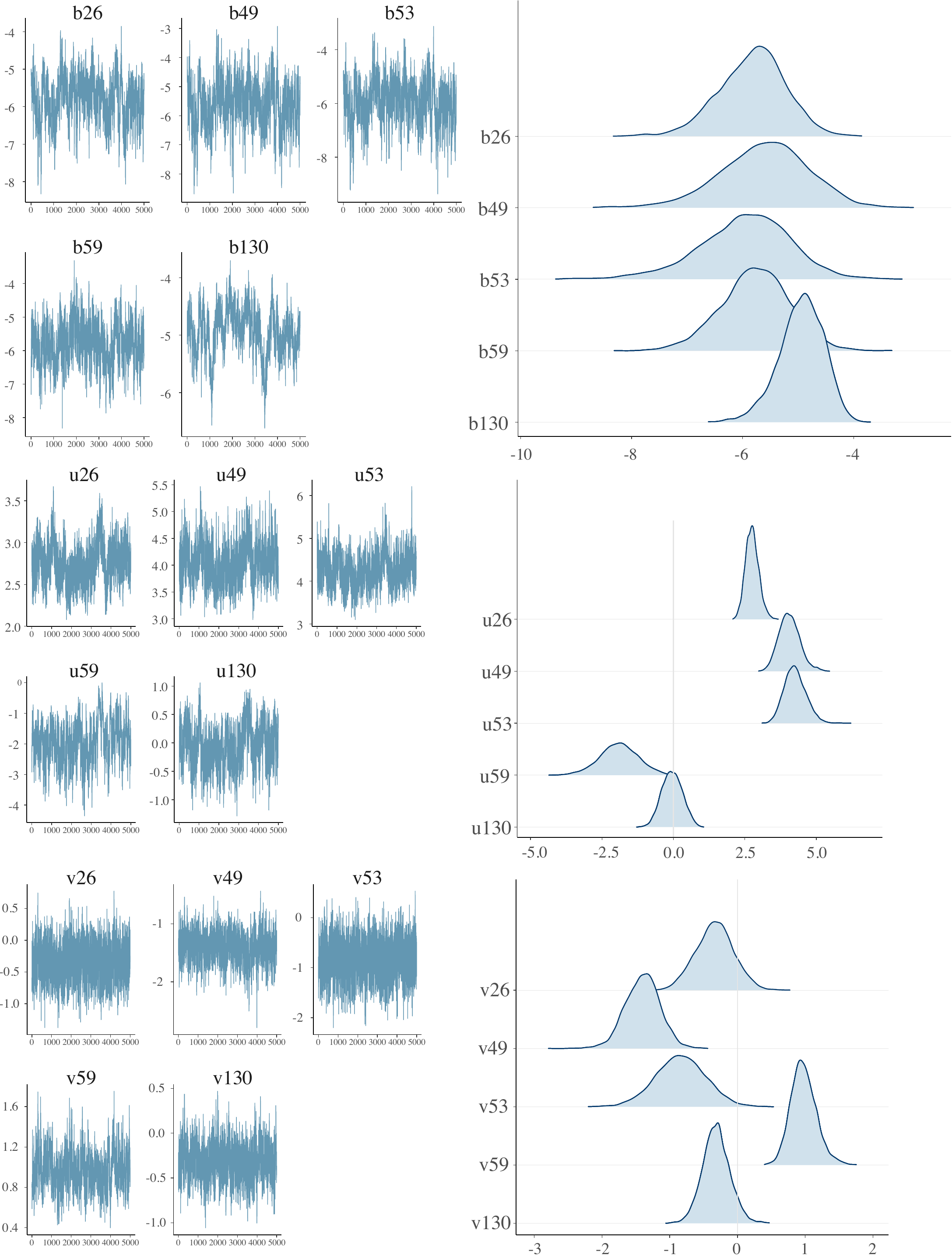}}
\caption{Trace plots and posteriors distributions of the popularity parameters, $b_a$, and latent variables, $u_a$ and $v_a$, of the five actors $a= 26$, 49, 53, 59, and 130 in the subset of the Enron network.}
\label{fig_subset_posteriors_q1}
\end{figure}

Figure \ref{fig_subset_posteriors_q1} shows the trace plots of the posteriors of the popularity parameters and latent variables of the actors for the model with a 1-dimensional latent variable for the actors and messages (the other models showed comparable results). Overall the plots show reasonably good mixing based on 500,000 draws of which every 100th draw is stored. Moreover we see that the posteriors of the latent variables for the actors as receivers ($u_a$) differ from the latent variables as sender ($v_a$), which suggests that actors have different roles as sender and receiver regarding the unobserved heterogeneity. 

\begin{figure}[t]
\centering
\makebox{\includegraphics[width=13.0cm]{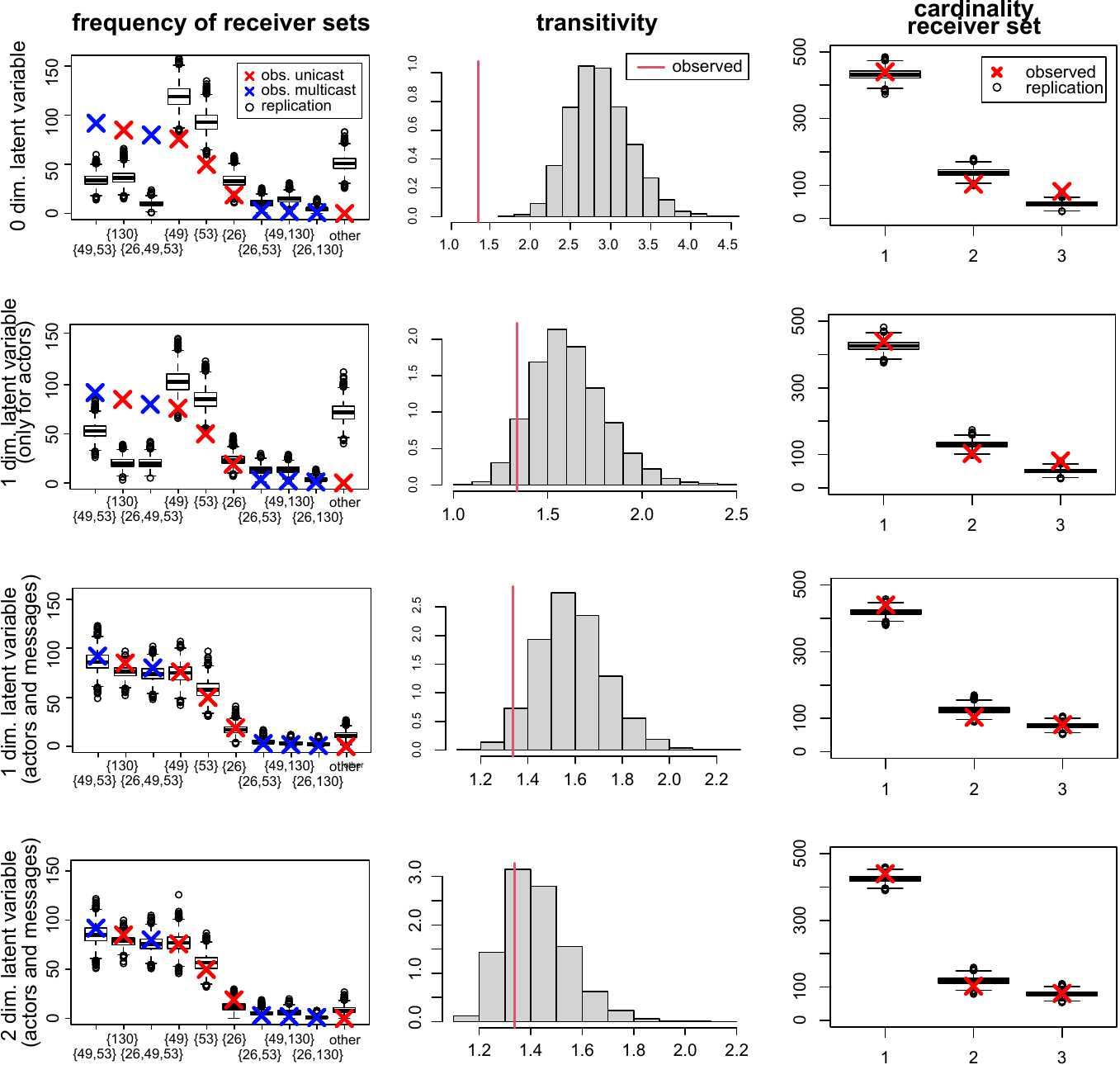}}
\caption{Posterior predictive distributions of the frequencies of the receiver sets of messages sent by actor 59 (left column), transitivity in the data (middle column), and the cardinality of the receiver set (right column) for a model with no latent variables (first row), a model with 1 latent variable with no latent variables for the messages (second row), a model with 1 latent variable for all actors and messages (third row), and a model with a 2 dimensional latent variable for all actors and messages (fourth row). The crosses and vertical lines correspond to the respective quantities in the observed data.}
\label{fig_subset}
\end{figure}

Figure \ref{fig_subset} (left column) shows the replicated frequencies for the nine observed receiver sets of messages sent by actor 59 and the remaining possible receivers sets (denoted by `other') using the four fitted models. We see that the first model with no latent variables and the second model with only a 1-dimensional latent variable for the actors and not for the messages both badly fit the observed data, while the latter two models with latent variables for both the actors and messages fit the data quite well. Figure \ref{fig_subset} (middle panels) shows the posterior predictive distributions of the first transitivity statistic for the four models (the plots for the second statistic looked comparable). We see that only the mc-amen model with no latent variables results in a poor fit while the other models fit this property of the data quite well, in particular the model with a 2 dimensional latent variable. Figure \ref{fig_subset} (right column) shows that all four models capture the cardinality of the receiver sets over all messages in the data quite well. This again illustrates that the mc-amen model (also without any latent variables) is able to capture the observed distribution of the number of receivers per message in the complete data. Based on all these results, the third mc-amen model with a 1-dimensional latent variable for the actors and messages seems to be preferred as it results in an accurate fit to the data and the it is more parsimonious than the fourth model with a 2-dimensional latent variable.

\begin{figure}[t]
\centering
\makebox{\includegraphics[width=8.0cm]{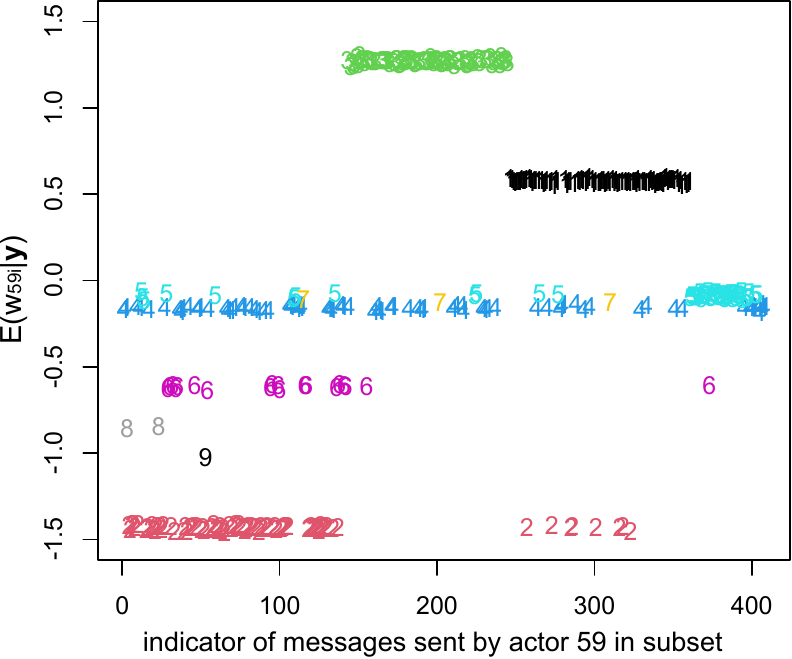}}
\caption{Posterior means of the latent variable of the 408 messages sent by actor 59 to actors 26, 49, 53, and 130. The numbers refer to the observed receiver sets of this sender from Figure \ref{fig_subset}: `1' = \{49,53\}, `2' = \{130\}, `3' = \{26,49,53\}, `4' = \{49\}, `5' = \{53\}, `6' = \{26\}, `7' = \{26,53\}, `8' = \{49,130\}, and `9' = \{26,130\}.}
\label{fig_postmean_W_q1}
\end{figure}

To give some insights of the importance of the latent variables of the messages to model multicast interaction behavior, Figure \ref{fig_postmean_W_q1} displays the posterior means of the latent variables of the 408 messages sent by actor 59 where the 9 different numbers (with different colors) refer to the 9 different observed receiver sets for sender 59 (see left panels in Figure \ref{fig_subset}), e.g., `1' refers to receiver set \{49,53\}, `2' refers to receiver set \{130\}, etc. As can be seen the posterior means of the latent variables are practically equal for the same receiver sets. This shows that the latent variables of the messages play an important role when fitting the observed receiver sets. It is also interesting to see that for second receiver set `2', i.e., \{130\}, the estimated latent variables are negative causing a large decrease of the suitability scores of actors 26, 49, and 53 due to their large and positive latent variables as receiver, i.e., $u_{26}$, $u_{49}$, and $u_{53}$ (Figure \ref{fig_subset_posteriors_q1}), and thus making only actor 130, with an estimated latent variable of approximately 0, to be a very suitable receiver. Similar explanations can be provided for the other receiver sets based on these results.

\subsection{Analysis of the full Enron email data}
The full Enron email data consisting of 21,635 messages which were sent among 156 actors between November 13, 1998, and June 21, 2002. The employees’ genders (male, female), seniority (junior, senior), their type of work (`Legal', `Trading', Other), title (`Specialist', `Administrator', etc.) and their department (ENA Gas Financial, Energy Operations, etc.) were also available. Following we Perry \& Wolfe (2013) we considered 7 different inertia and reciprocity statistics for 7 different time intervals, dummy covariates for all 16 combinations of senders and receivers working in `Legel' (1=yes, 0=no), working in `Trading' (1=yes, 0=no), being junior (1=yes, 0=no), being female (1=yes, 0=no). We also added two dummy covariates of whether sender and receiver have the same title and whether they work in the same department. In total this resulted in 32 covariates. All these covariates were standardized over the entire dataset.

\begin{figure}[t]
\centering
\makebox{\includegraphics[width=12cm]{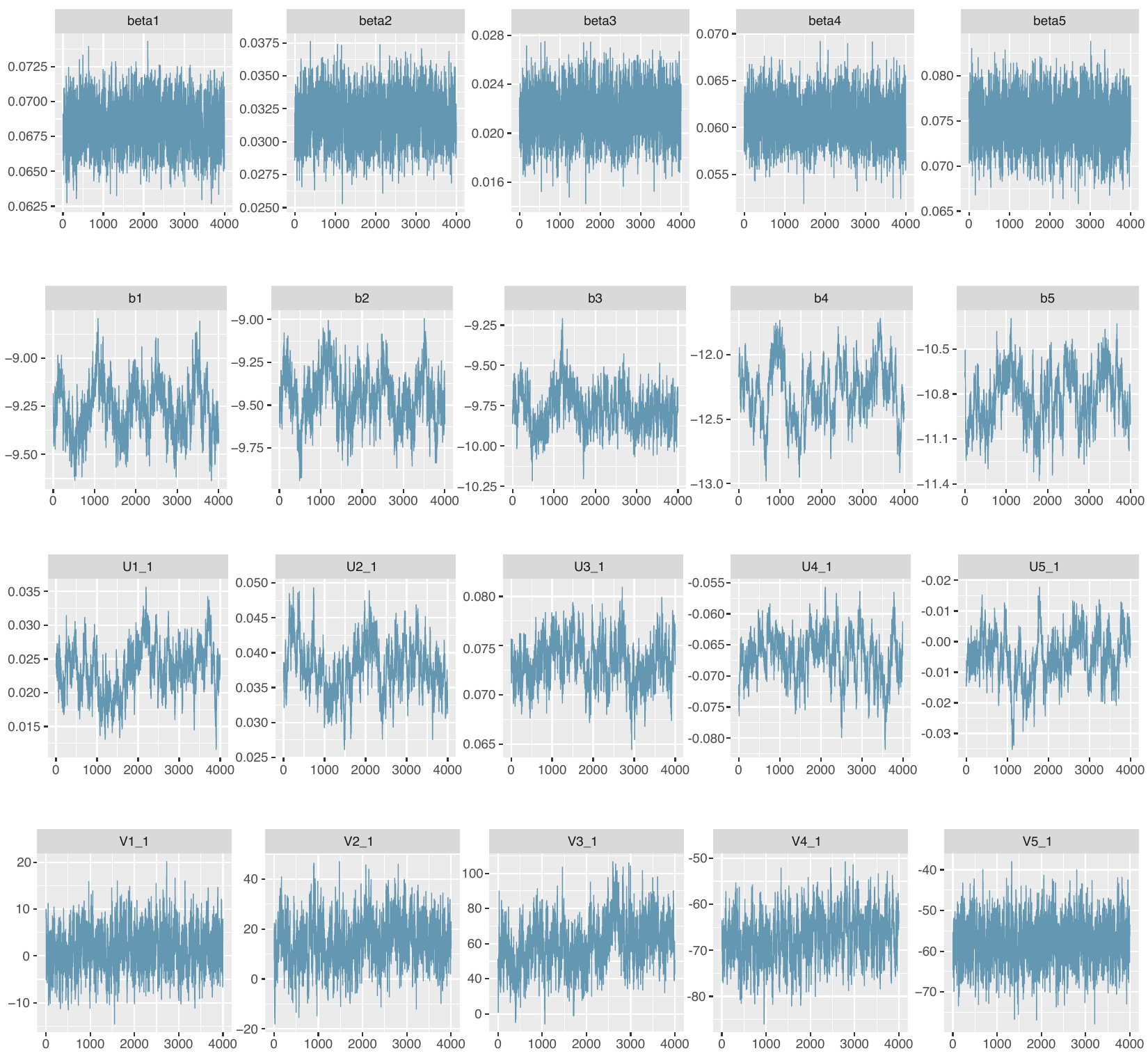}}
\caption{Trace plots of the first five coefficients (first row), the popularity parameters of the first five actors (second row), the first latent variable of the first five actors as receiver (third row), and the first latent variable of the first five actors as sender (fourth row).}
\label{fig_enron_draws}
\end{figure}

A mc-amen was fit to these data with $Q=0$, 1, 2, 3, and 4 dimensions. In the MCMC sampler, every 100th draw was stored. After burn-in, the trace plot of the posterior draws of the first five coefficients, the popularity parameters of the first five actors, the first latent variable of the first five actors as receiver, and the first latent variable of the first five actors as sender in the model with $Q=2$ dimensional latent variable are displayed in Figure \ref{fig_enron_draws}. The trace plots of the other parameters and other models were comparable and omitted to keep the results concise. Overall the mixing looks acceptable.

\begin{figure}[t]
\centering
\makebox{\includegraphics[width=12cm]{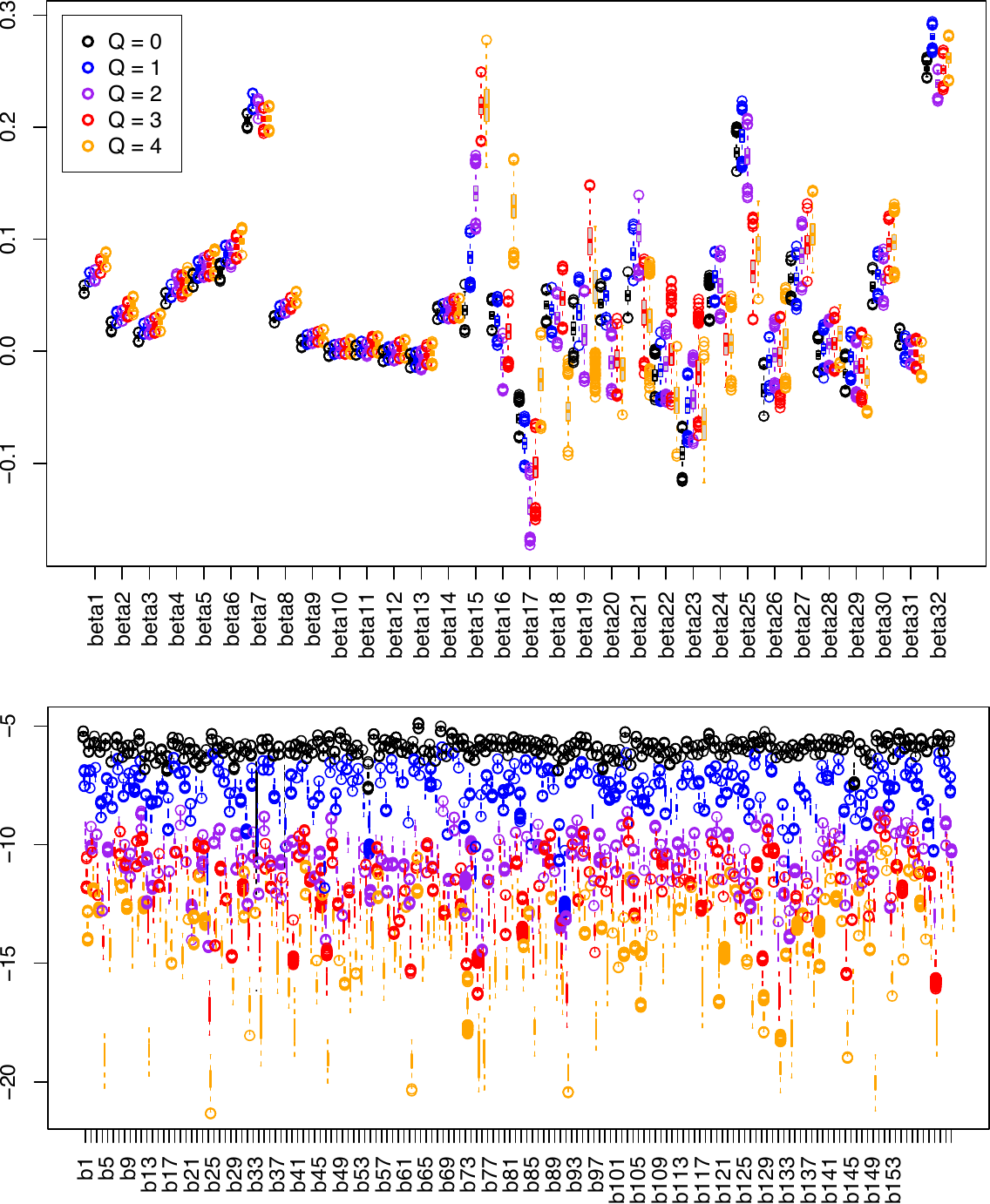}}
\caption{Posterior credibility intervals of the coefficients (upper panel) and the popularity parameter (lower panel) for a mc-amen model with a latent variable with $Q=0$, 1, 2, 3, and 4 dimensions fitted to the Enron email data.}
\label{fig_enron_beta_bvec}
\end{figure}

Figure \ref{fig_enron_beta_bvec} shows the posterior credibility intervals of the coefficients (first row) and the popularity parameter (second row) for a mc-amen model with a latent variable with $Q=0$, 1, 2, 3, and 4 dimensions fitted to the Enron email data. Again, we see that the relative size of the estimated coefficients is fairly consistent over all 32 parameters for the 5 different models with differing latent dimensionality. This suggests that conclusions about the relative importance of the covariates are largely unaffected of the dimensionality of the latent variable of the fitted mc-amen model. Furthermore, we see that the posterior intervals of the popularity parameters become lower on average as the dimensionality increases, similar as in the simulated data.
 
Next Figure \ref{fig_enron_pred} displays the posterior predictive distributions of the number of messages for the 20 most frequently observed combinations of senders and receivers in the observed data, the 21 to 80 most frequently observed combinations, the cardinality of the receiver sets over all data, and the two transitivity measures. In the first two columns the red and blue crosses denote the observed frequencies of the unicast and multicast messages, respectively. The model without latent variables (Figure \ref{fig_enron_pred}, first row) clearly shows the worst fit. The model badly captures the frequency even of the most often observed messages (from actor 20 to actor 131), and also badly captures the frequencies of certain multicast messages (specifically of actor 20 to actors $\{61,122,131\}$, and of actor 20 to actors $\{61,113,122,131\}$). The misfit of the frequency of the number of messages of actor 20 to actor 131 can be explained by the fact that 131 is a popular receiver of messages sent by actor 20 for both unicast messages as well as multicast messages. Because the model assumes that unicast messages are on average more likely than multicast messages (because this is suggested from the data), the predicted frequencies of unicast messages of actor 20 to actor 131 become overestimated. This misfit considerably decreases for the mc-amen with a 1-dimensional latent variable (Figure \ref{fig_enron_pred}, second row) even though the frequencies of some of these 20 most commonly observed messages still show some misfit. This misfit has largely disappeared for the mc-amen model with a 2-dimensional latent variable (Figure \ref{fig_enron_pred}, third row). For the 20 most frequently observed messages, the fit does not become much better for the mc-amen with a 3-dimensional latent variable (Figure \ref{fig_enron_pred}, fourth row) or a 4-dimensional latent variable (Figure \ref{fig_enron_pred}, fifth row). The same conclusion can be drawn when looking at the cardinality of the receiver set in the complete data set (Figure \ref{fig_enron_pred}, third column) and the first transitivity measure (Figure \ref{fig_enron_pred}, fourth column). For the second transitivity measure, the misfit may still be considered too large for the mc-amen with 2 dimensions. Based on this measure, a mc-amen model with 3 or 4 dimensions may be preferred. To avoid an overly complex model while still maintaining a reasonable fit, the mc-amen model with a 2-dimensional latent variable may be preferred based on these posterior predictive checks.

\begin{figure}[t]
\centering
\makebox{\includegraphics[width=14.5cm]{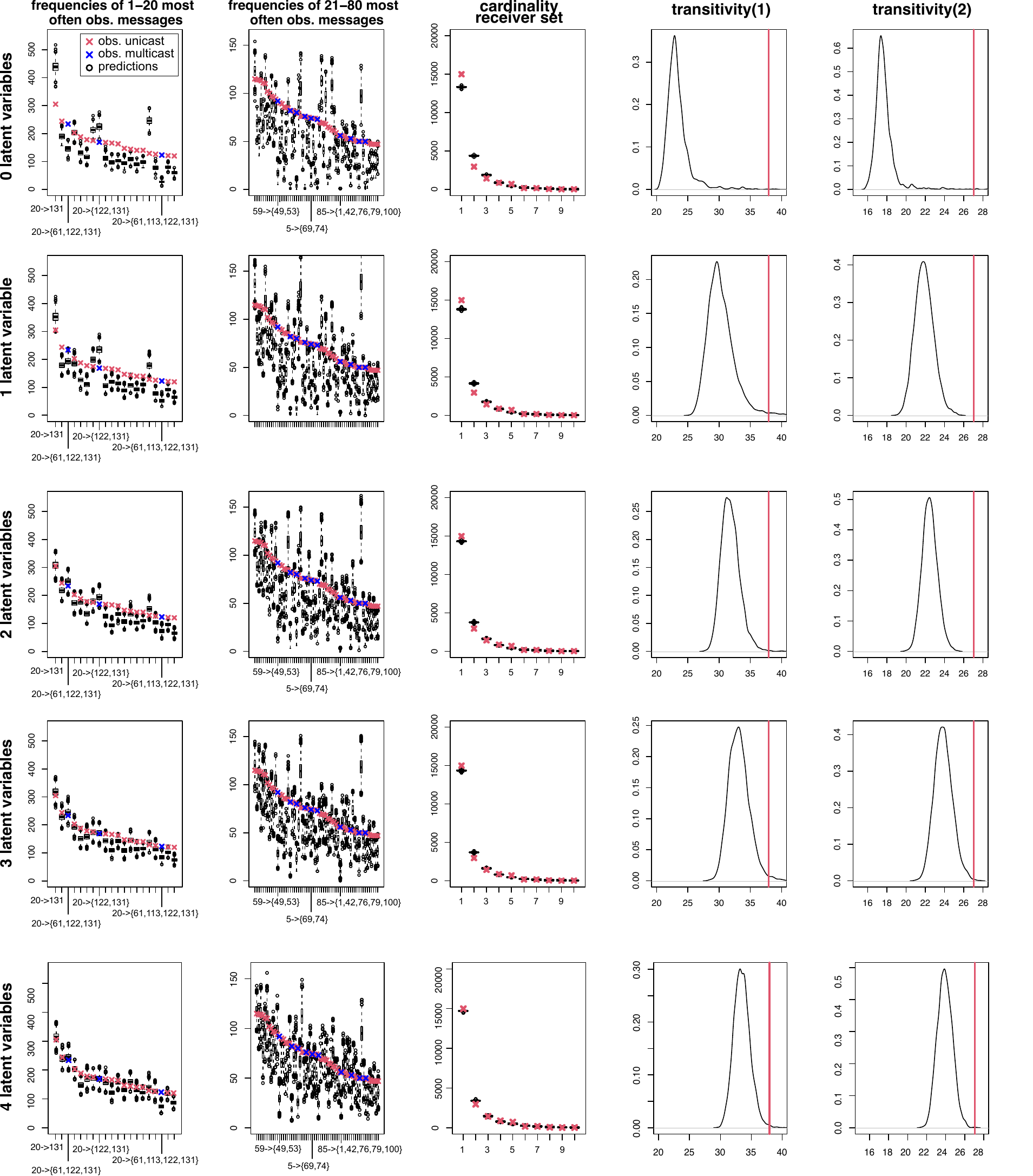}}
\caption{Posterior predictive distributions of the frequencies of the 1 to 20 most frequently observed messages (first column), the frequencies of the 21 to 80 most frequently observed messages (second column), the distribution of the cardinality of the receiver set over all messages (third column), the two transitivity measures (fourth and fifth column) for a model with 0, 1, 2, 3, and 4 dimensions of the latent variable (first to fifth row). The crosses and vertical lines correspond to the respective quantities in the observed data.}
\label{fig_enron_pred}
\end{figure}

\begin{figure}[t]
\centering
\makebox{\includegraphics[width=14cm]{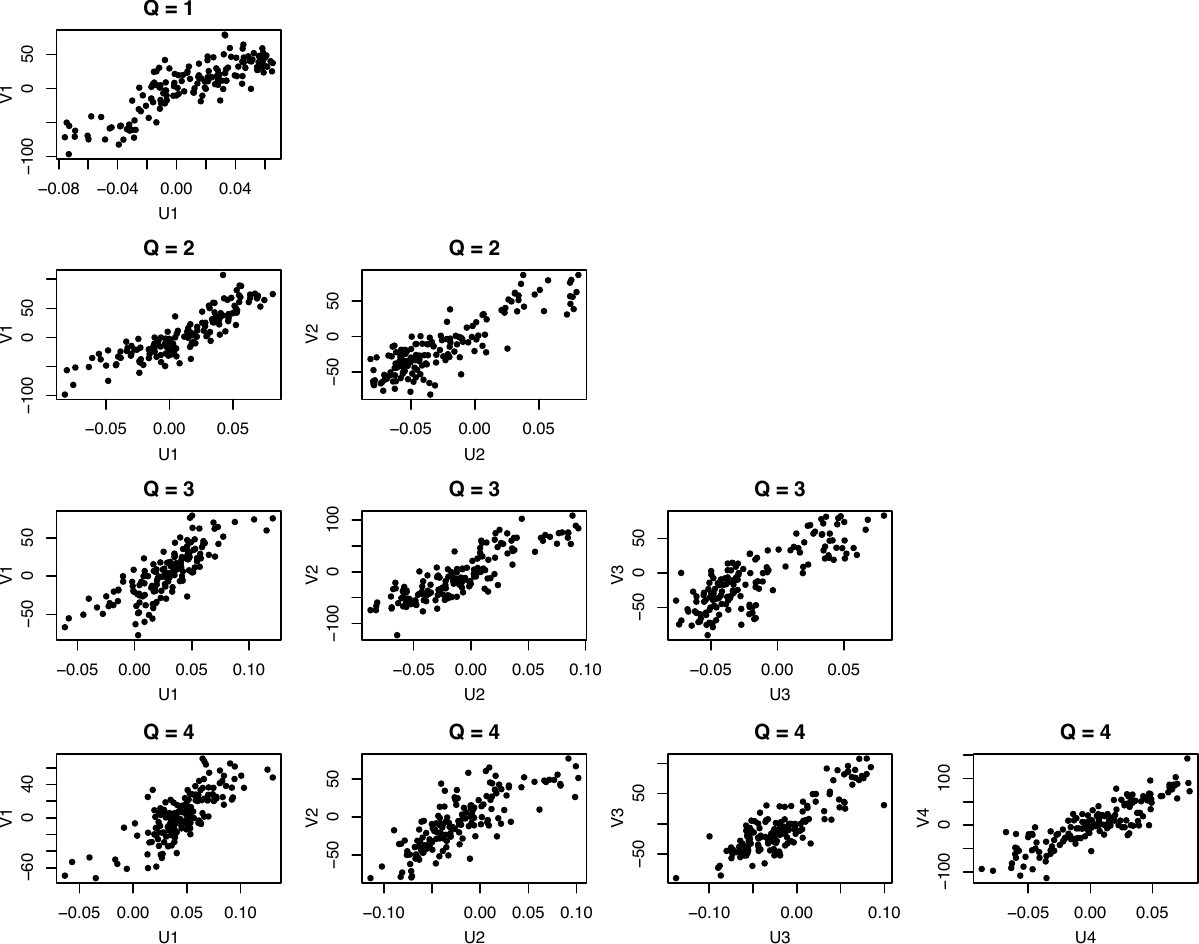}}
\caption{Posterior means of each latent variable of an actor as receiver ($u_a$) versus the latent variable as sender ($v_a$) for all latent variables in a mc-amen model with $Q=1$, 2, 3, and 4 dimensions for the latent variable.}
\label{fig_enron_uv}
\end{figure}

Next we investigate how strong the latent variables of the actors as sender and as receiver are correlated. For this purpose the posterior means of latent variables under all models are computed. The results are displayed in Figure \ref{fig_enron_uv}. The plots show the anticipated positive correlations for all latent variables under all models. Moreover we see that the variability of the latent variables of actors as sender is considerably larger than the variability of the latent variable of actors as receiver. This however does not imply that actors vary more in their tendency to send messages in comparison to their tendency to receive messages because the fit is invariant to the scale of the latent variables, i.e., $\textbf{u}_{a'}^{\top}\textbf{v}_{a}=(h\textbf{u}_{a'})^{\top}(h^{-1}\textbf{v}_{a})$ and $\textbf{u}_{a'}^{\top}\textbf{w}_{ai}=(h\textbf{u}_{a'})^{\top}(h^{-1}\textbf{w}_{ai})$, for any constant $h>0$. In the fitted model the standard deviations of the estimated latent variables of the messages were approximately 30, which is comparable to the variability of the latent sender variables.

Based on these results, we can conclude the following. First, the mc-amen model is able to capture the (typical) decreasing shape of the empirical distribution of the cardinality of the receiver set in the Enron data (Figure \ref{hist_obs_m}) quite well. Second, the multiplicative latent factor structure in the mc-amen model is able to capture the multicast interaction behavior between actors quite well. Third, increasing the dimensionality of the latent variables generally result in an improved fit to the data. To avoid an overly complex model for the Enron data, the mc-amen model with a 2-dimensional latent variable would be preferred. Fourth, the relative magnitude of the coefficients (fixed effects) depends little on the dimensionality of the latent variables.

\section{Discussion}
A multicast additive and multiplicative effects network (mc-amen) model was proposed for analyzing relational data that contain unicast messages (i.e., an observation of one actor towards another actor) and multicast messages (i.e., an observation of one actor towards multiple other actors). The multicast behavior is captured by placing the potential receivers of a given message sent by a given receiver on a suitability scale where a larger suitability score of a potential receiver results in a larger probability of being included in the receiver set. The suitability score is modeled using probit regression model of observed predictors variables and latent variables. Subsequently a threshold parameter is drawn from a truncated normal distribution with an upper bound that is equal to the largest suitability score. Actors are included in the receiver set who have a larger suitability score than the threshold parameter. To capture any unobserved heterogeneity regarding the social interaction behavior in the network, a multiplicative approach is considered for the latent variables. This multiplicative approach encompasses both the discrete stochastic block model as well as the continuous latent distance model (\cite{Hoff:2008}). The mc-amen includes separate latent variables for the actors as sender and as potential receiver. Thereby, actors are allowed to have a different (latent) tendency to send messages as to receive messages. Furthermore, by including latent variables for the individual messages, it is possible to capture heterogeneous interaction behavior between a given sender and potential receivers.

The analyses of the Enron email data showed that the mc-amen model can accurately capture the empirical distribution of the cardinality of receiver sets that is typically observed in email data (where the probability tends to decrease as the cardinality increases, making unicast messages more likely than messages to two receivers, which in turn is more likely than messages to three receivers, etc.). Furthermore, the model is able to capture heterogeneous interaction behavior between actors in unicast messages as well as multicast messages. Furthermore, the data showed a positive correlation between the latent variables of the actors as receivers and as senders, which suggests that actors have a similar (but not identical) interaction behavior as a sender and as a receiver. To assess model fit posterior predictive checks were use based on the observed frequencies of the most commonly observed combinations of senders and receivers, the transitivity in the data, and the distribution of the cardinality of the receiver sets. This resulted in preference for a mc-amen model with a 2 dimensional latent variable for the Enron data (where a more parsimonious model with fewer latent variables was preferred to avoid an overly complex model).

Different directions would be important to pursue for future research. In particular it would be important to improve the computational efficiency of the algorithm when fitting the model. Because the mc-amen has many latent variables which depend on each other in a complex manner, posterior mixing can be slow resulting large autocorrelations (for this reason we only used every 100th posterior draw in our analyses). Mixing could potentially be improved using orthogonal transformations of the parameters and by sampling the parameters in larger blocks where possible. Another potential computational improvement could be achieved using variational Bayesian algorithms. Another interesting direction for future research is to explore other distributions of the threshold parameters than the truncated normal distributions. This is important for data with other types of shapes for the empirical distributions of the cardinality of the receiver set, for instance when multicast messages are more frequently observed than unicast messages. A last interesting direction for future research we mention here is to extend the mc-amen model with an additional clustering step which will allocate the actors in the same clusters if they show similar interaction behavior \cite[e.g., similar as was done by][ for the latent distance model]{handcock2007model}. This could be done using a multivariate normal mixture model for the latent variables. Note that even though the current mc-amen model does not parameterize this clustering step explicitly, it would of course be possible to identify clusters of actors if their estimated latent variables are close to each other based on the fitted mc-amen model.
\\

\noindent \textbf{Acknowledgements.} The authors would like to thank two anonymous reviewers and an associate editor for their constructive feedback which resulted in important improvements of the manuscript. The first author was supported by an ERC Starting Grant (758791).

\bibliographystyle{apacite}
\bibliography{refs_mulder}

\appendix

\section{Figures of Simulation 1 (Section \ref{simulations1})}

\begin{figure}[h]
\centering
\makebox{\includegraphics[width=14cm]{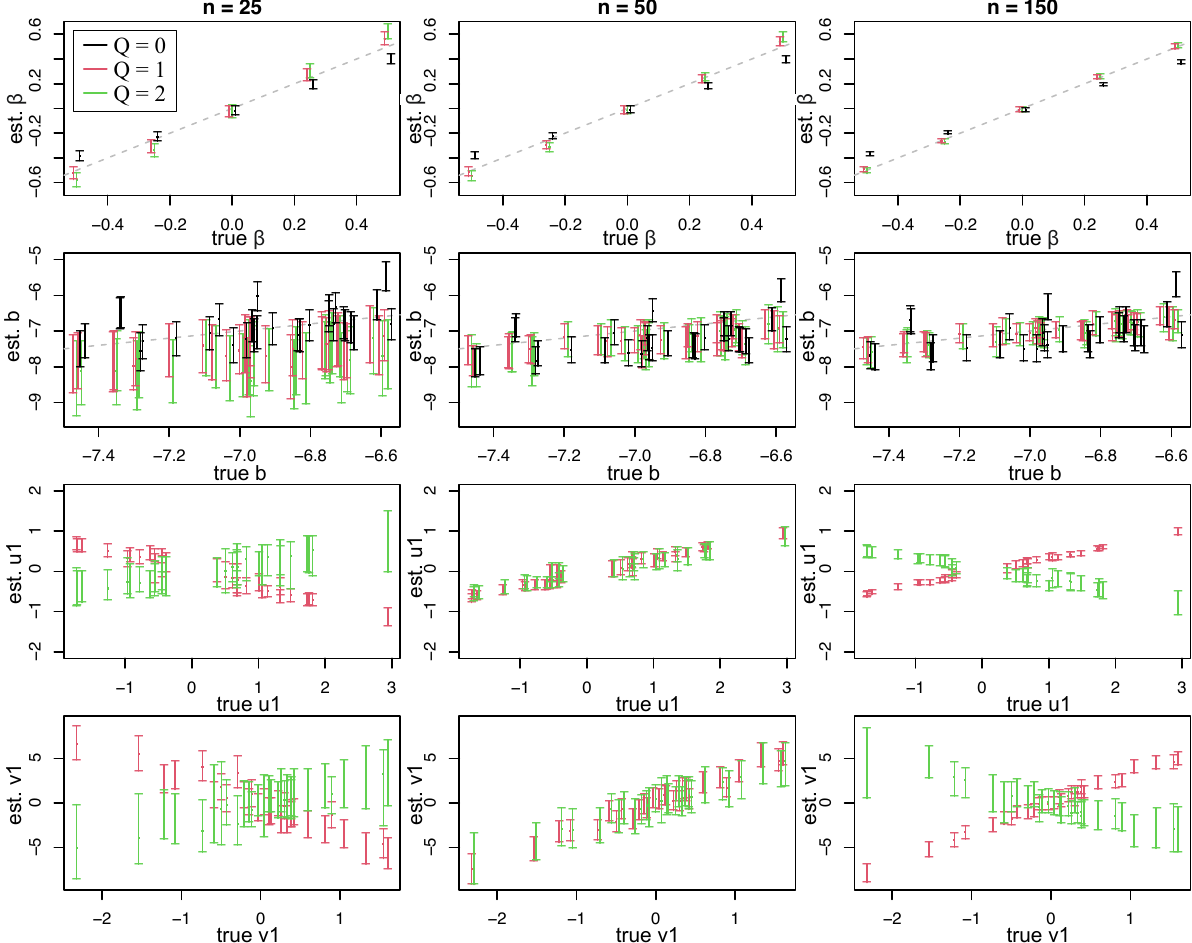}}
\caption{Population 0 (Table \ref{tabExtraSim}). 95\% credibility intervals for the fixed effects, the random popularity effects, and the latent variables of the actors as senders and receivers in the case of 25 (left panels), 50 (middle panels), and 150 (right panels) messages per actor. The results for a model with 0, 1, and a 2 dimensional latent variable are displayed in black, red, and green, respectively.}
\label{CrI_an1}
\end{figure}

\begin{figure}[h]
\centering
\makebox{\includegraphics[width=14cm]{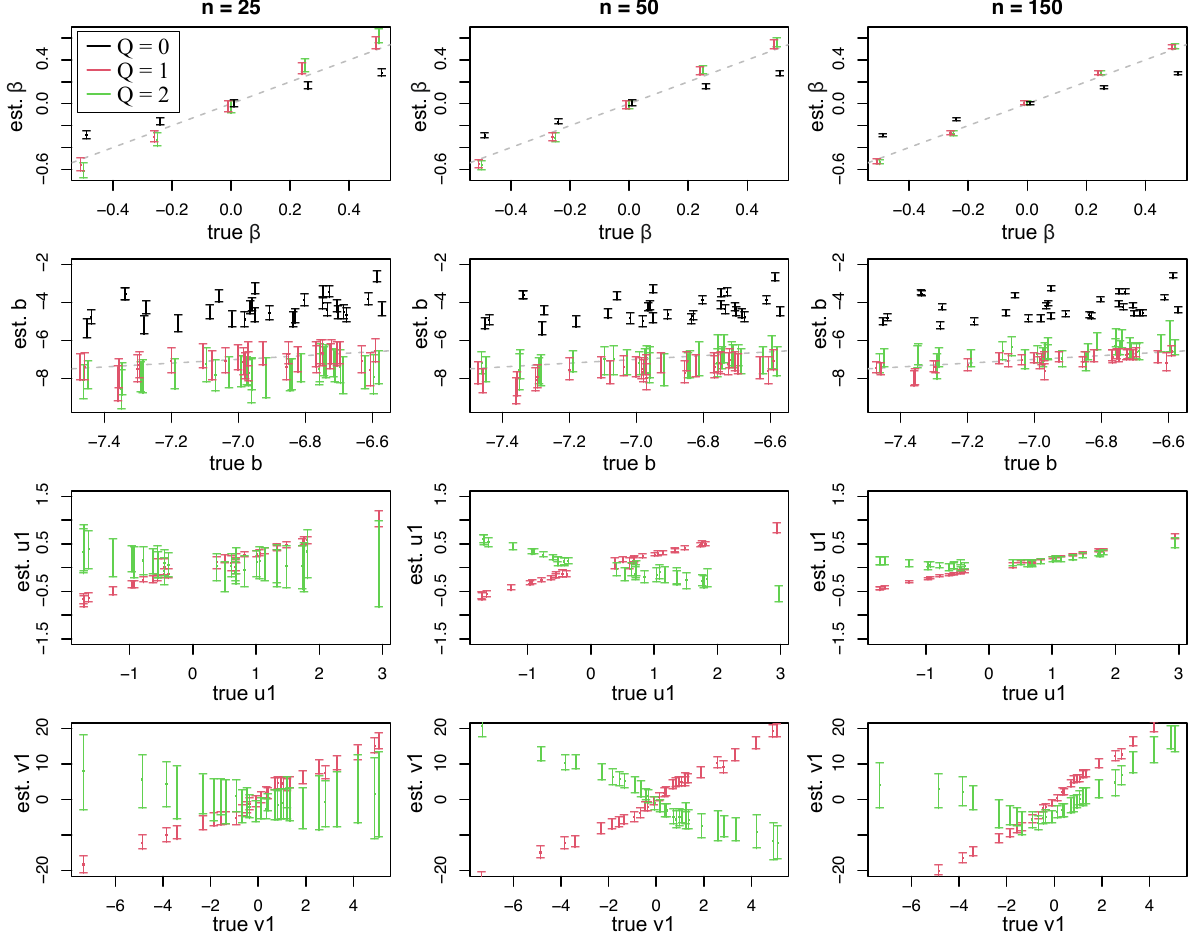}}
\caption{Population 1a (Table \ref{tabExtraSim}). 95\% credibility intervals for the fixed effects, the random popularity effects, and the latent variables of the actors as senders and receivers in the case of 25 (left panels), 50 (middle panels), and 150 (right panels) messages per actor. The results for a model with 0, 1, and a 2 dimensional latent variable are displayed in black, red, and green, respectively.}
\label{CrI_an2}
\end{figure}

\begin{figure}[h]
\centering
\makebox{\includegraphics[width=14cm]{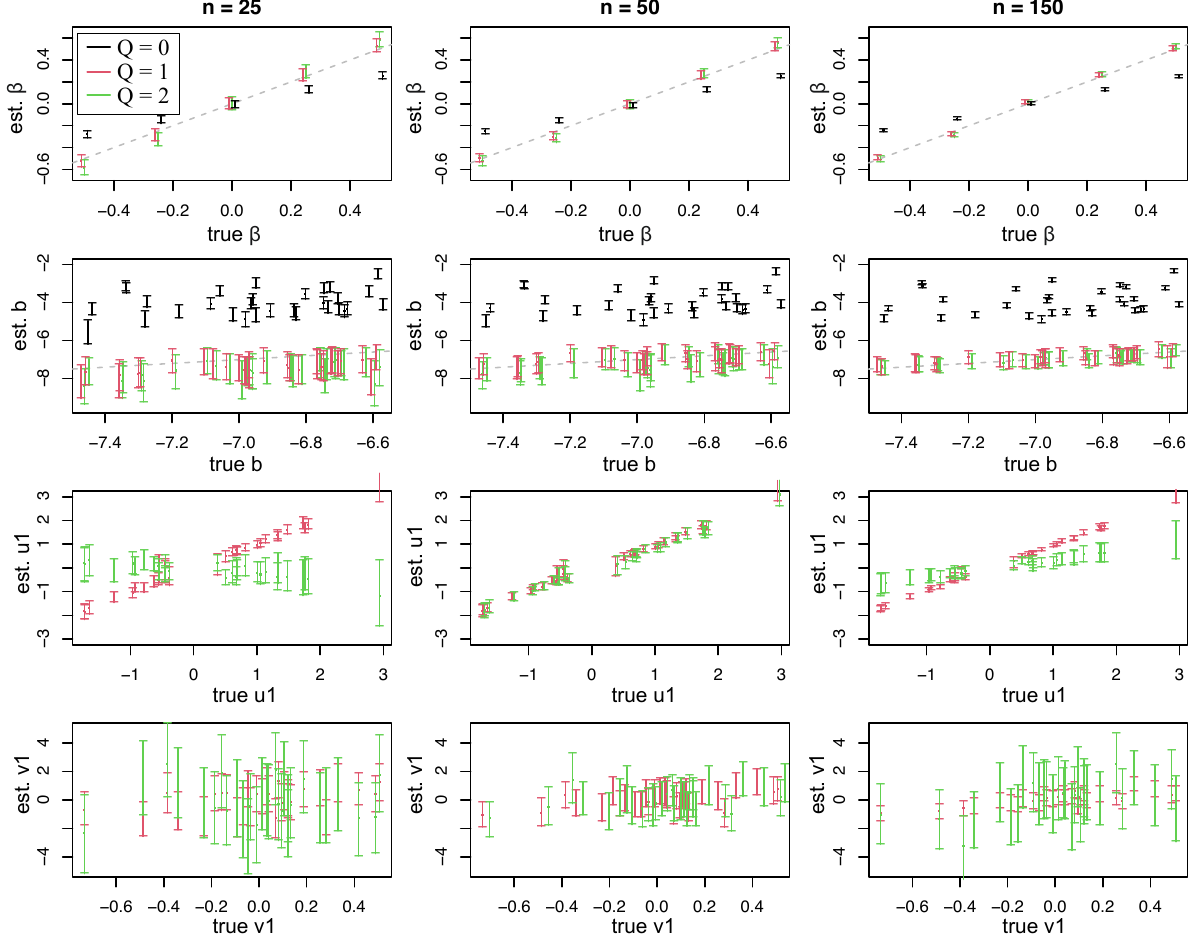}}
\caption{Population 1b (Table \ref{tabExtraSim}). 95\% credibility intervals for the fixed effects, the random popularity effects, and the latent variables of the actors as senders and receivers in the case of 25 (left panels), 50 (middle panels), and 150 (right panels) messages per actor. The results for a model with 0, 1, and a 2 dimensional latent variable are displayed in black, red, and green, respectively.}
\label{CrI_an3}
\end{figure}

\begin{figure}[h]
\centering
\makebox{\includegraphics[width=14cm]{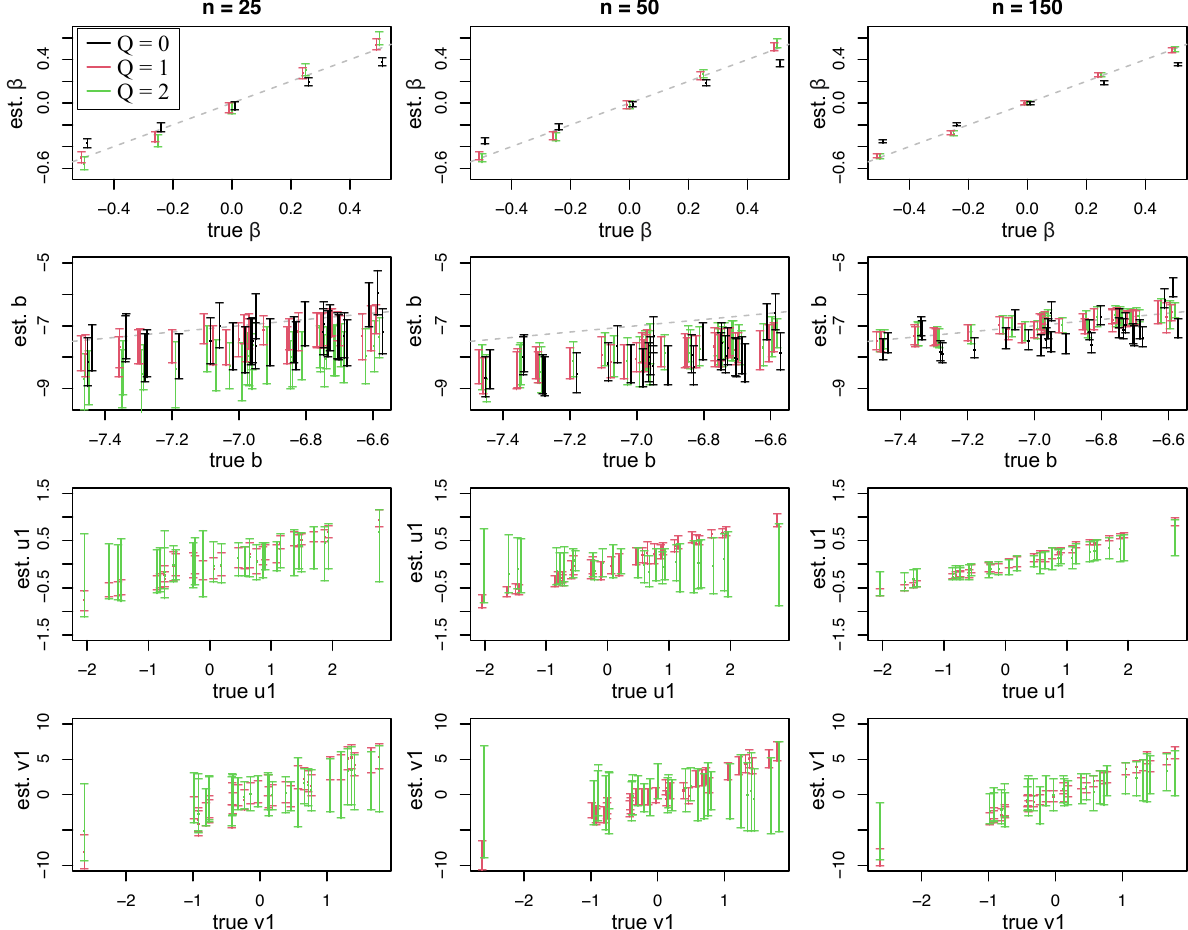}}
\caption{Population 2a (Table \ref{tabExtraSim}). 95\% credibility intervals for the fixed effects, the random popularity effects, and the latent variables of the actors as senders and receivers in the case of 25 (left panels), 50 (middle panels), and 150 (right panels) messages per actor. The results for a model with 0, 1, and a 2 dimensional latent variable are displayed in black, red, and green, respectively.}
\label{CrI_an4}
\end{figure}

\begin{figure}[h]
\centering
\makebox{\includegraphics[width=14cm]{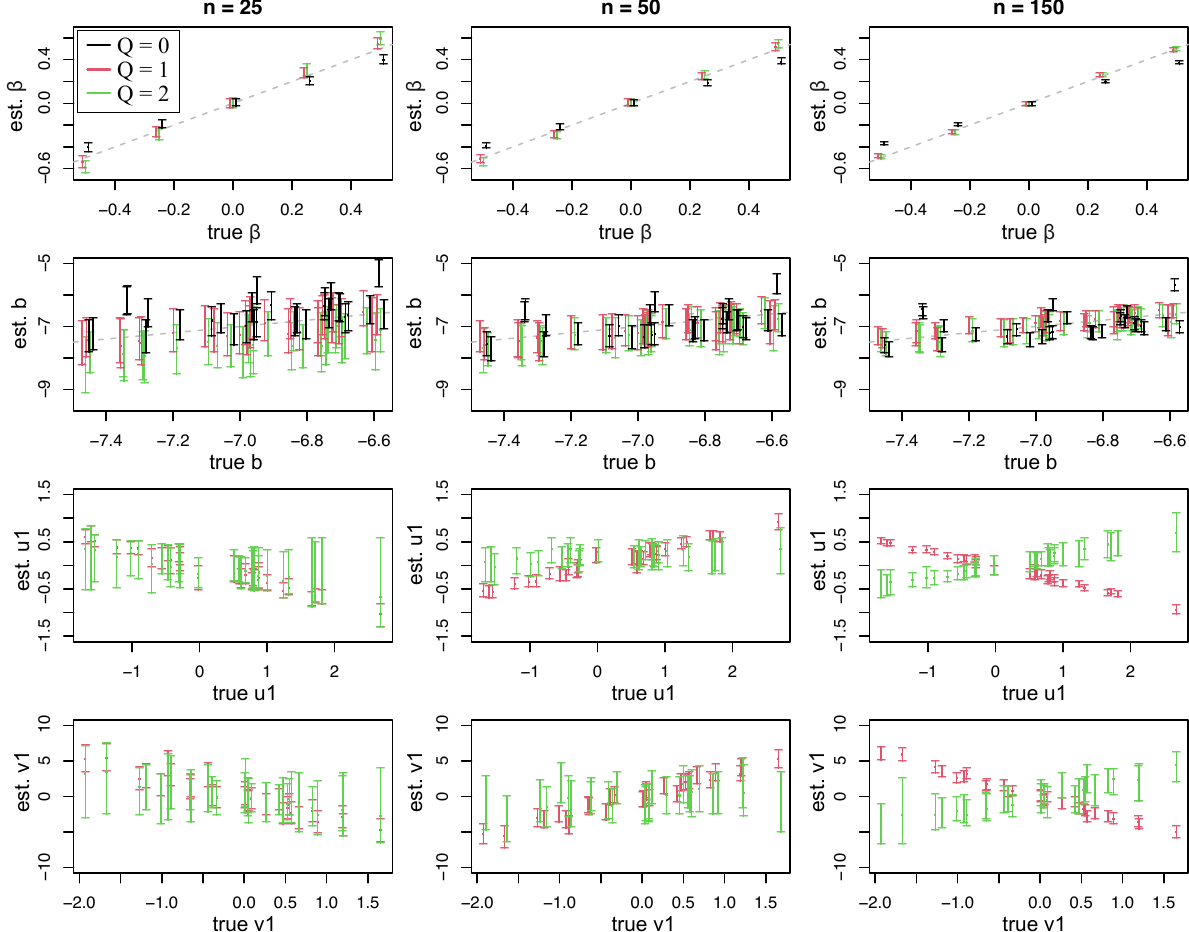}}
\caption{Population 2b (Table \ref{tabExtraSim}). 95\% credibility intervals for the fixed effects, the random popularity effects, and the latent variables of the actors as senders and receivers in the case of 25 (left panels), 50 (middle panels), and 150 (right panels) messages per actor. The results for a model with 0, 1, and a 2 dimensional latent variable are displayed in black, red, and green, respectively.}
\label{CrI_an5}
\end{figure}

\begin{figure}[h]
\centering
\makebox{\includegraphics[width=14cm]{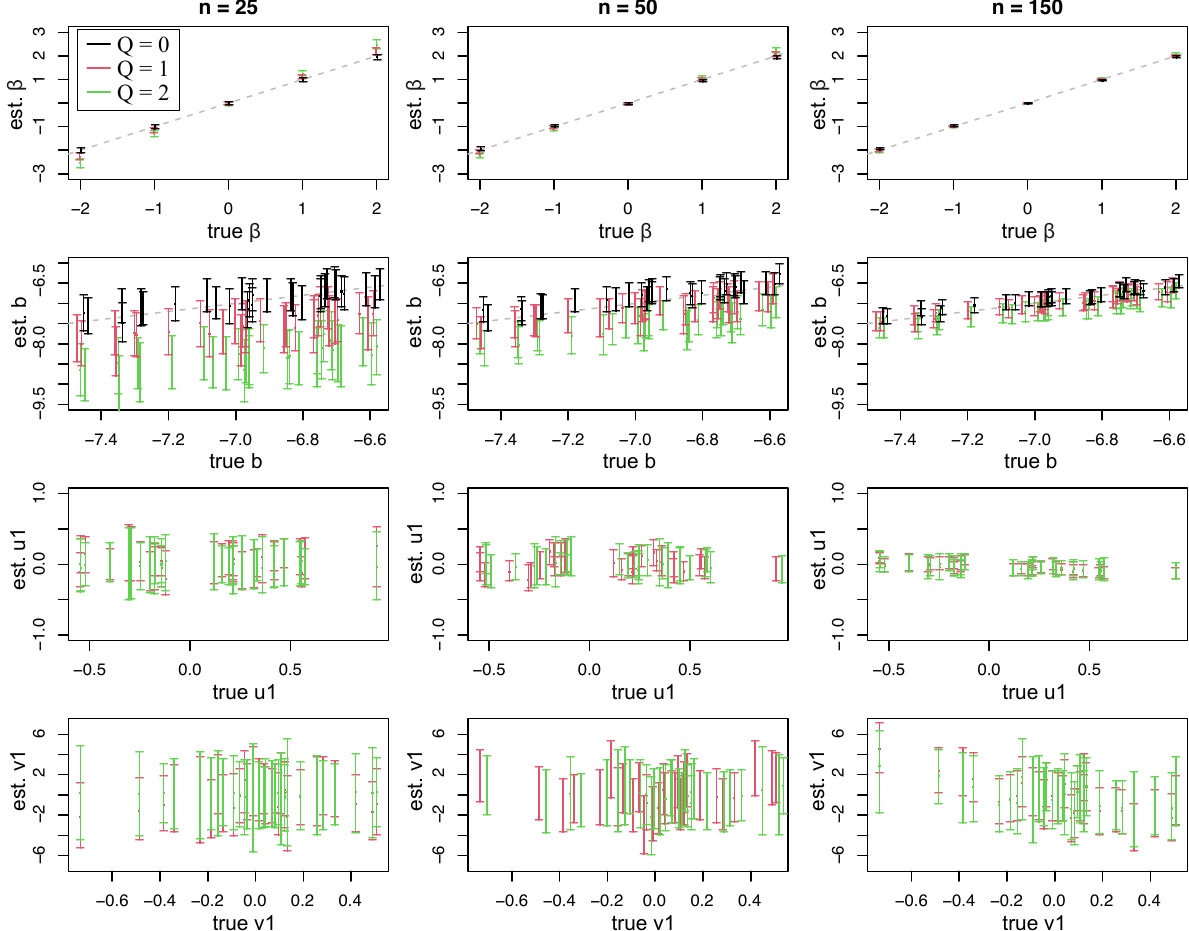}}
\caption{Population 3 (Table \ref{tabExtraSim}). 95\% credibility intervals for the fixed effects, the random popularity effects, and the latent variables of the actors as senders and receivers in the case of 25 (left panels), 50 (middle panels), and 150 (right panels) messages per actor. The results for a model with 0, 1, and a 2 dimensional latent variable are displayed in black, red, and green, respectively.}
\label{CrI_an6}
\end{figure}

\begin{figure}[h]
\centering
\makebox{\includegraphics[width=14cm]{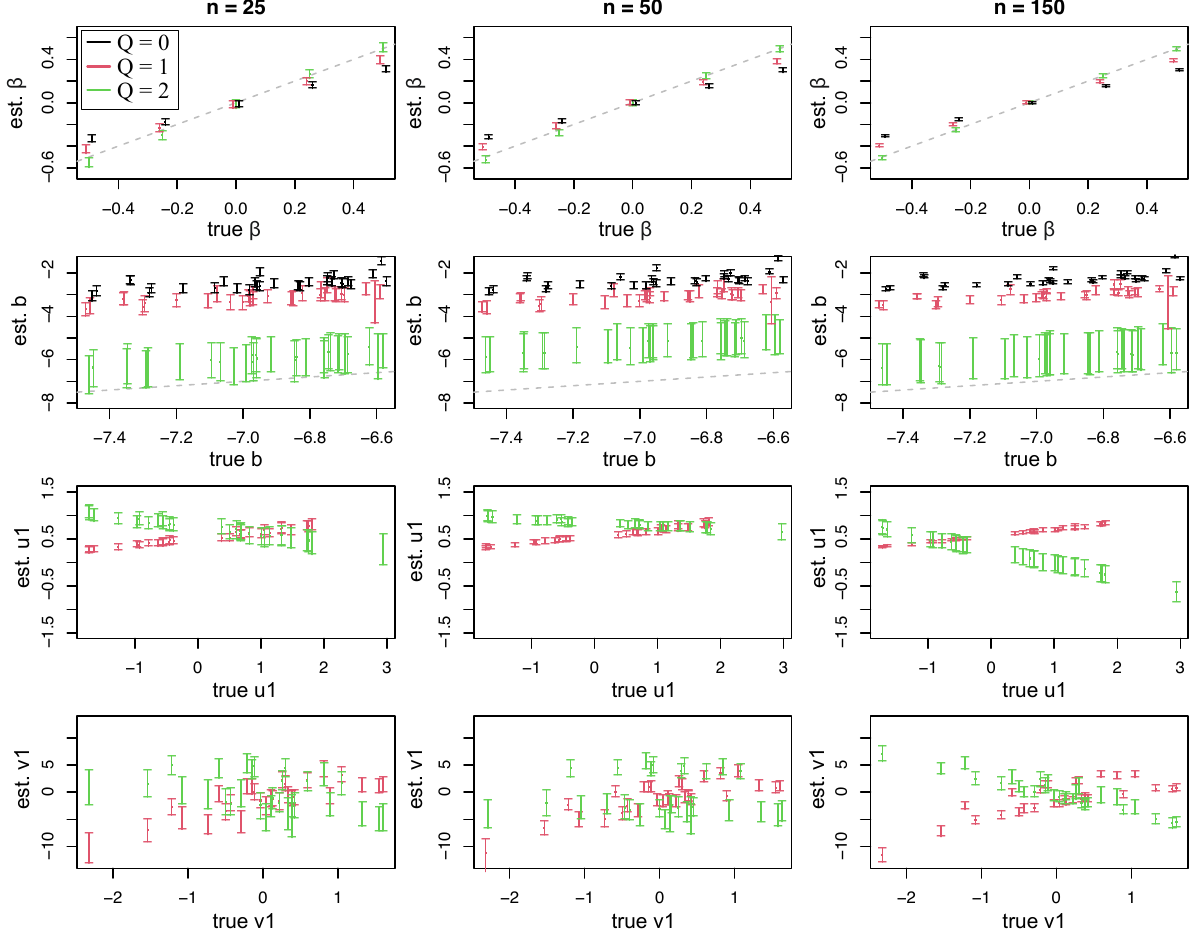}}
\caption{Population 4 (Table \ref{tabExtraSim}). 95\% credibility intervals for the fixed effects, the random popularity effects, and the latent variables of the actors as senders and receivers in the case of 25 (left panels), 50 (middle panels), and 150 (right panels) messages per actor. The results for a model with 0, 1, and a 2 dimensional latent variable are displayed in black, red, and green, respectively.}
\label{CrI_an7}
\end{figure}


\begin{figure}[h]
\centering
\makebox{\includegraphics[width=14cm]{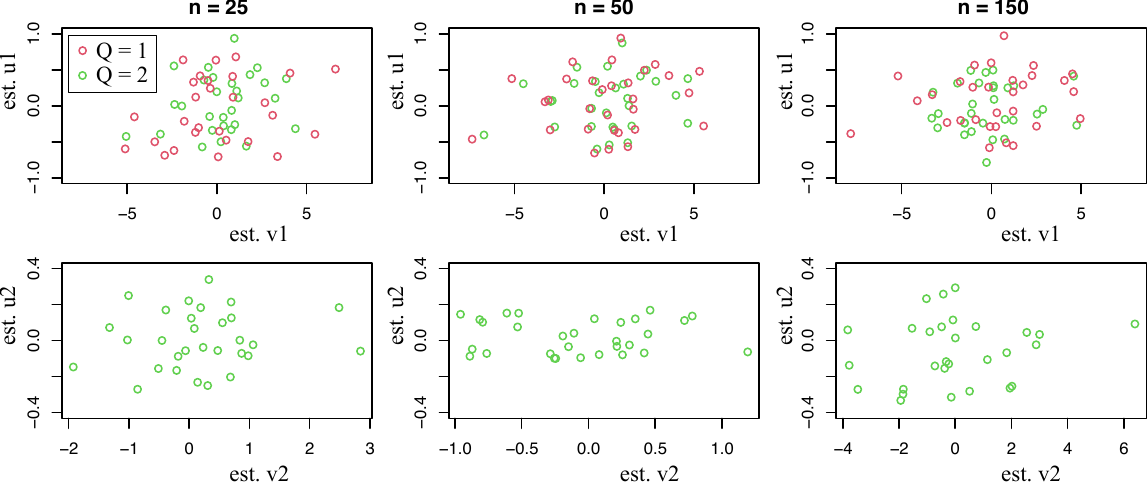}}
\caption{Population 0 (Table \ref{tabExtraSim}). Estimated latent variables of the actors as sender ($x$ axis) and as receiver ($y$ axis) for the first dimension (upper panels) and the second dimension (lower panels) in the case of 25 (left panels), 50 (middle panels), and 150 (right panels) messages per actor. The results for a model with a 1 and 2 dimensional latent variable are displayed in red and green, respectively.}
\label{vu_an1}
\end{figure}

\begin{figure}[h]
\centering
\makebox{\includegraphics[width=14cm]{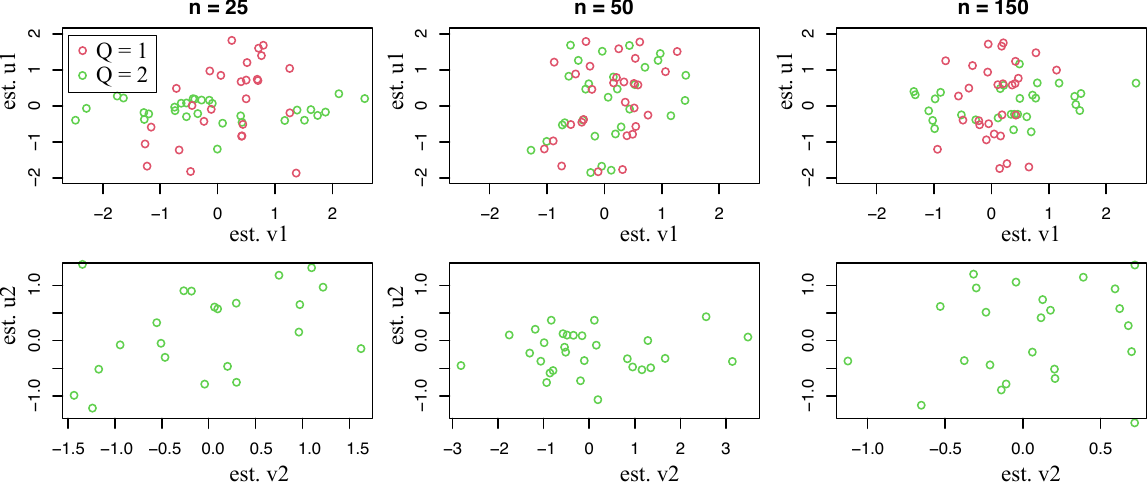}}
\caption{Population 1a (Table \ref{tabExtraSim}). Estimated latent variables of the actors as sender ($x$ axis) and as receiver ($y$ axis) for the first dimension (upper panels) and the second dimension (lower panels) in the case of 25 (left panels), 50 (middle panels), and 150 (right panels) messages per actor. The results for a model with a 1 and 2 dimensional latent variable are displayed in red and green, respectively.}
\label{vu_an2}
\end{figure}

\begin{figure}[h]
\centering
\makebox{\includegraphics[width=14cm]{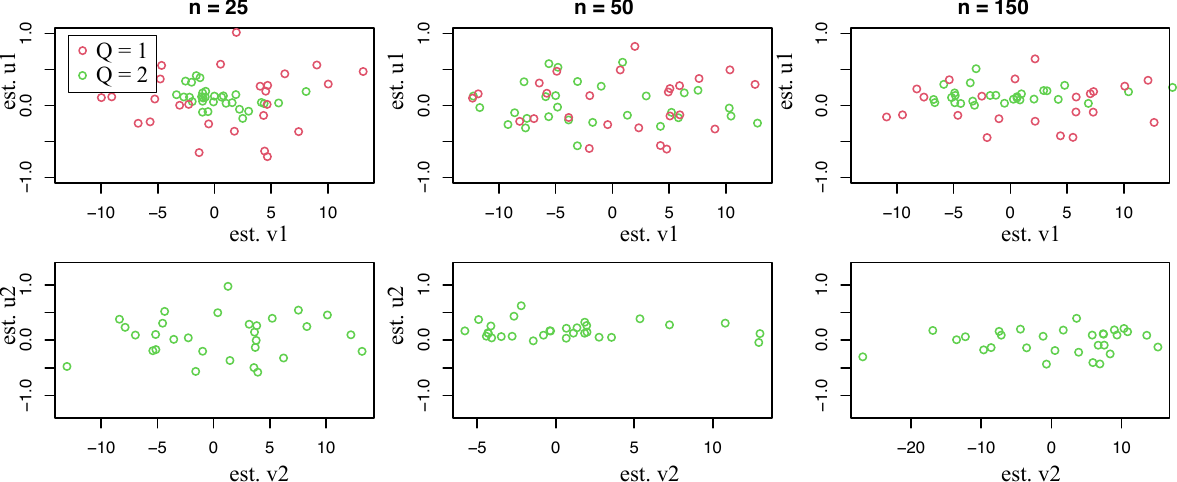}}
\caption{Population 1b (Table \ref{tabExtraSim}). Estimated latent variables of the actors as sender ($x$ axis) and as receiver ($y$ axis) for the first dimension (upper panels) and the second dimension (lower panels) in the case of 25 (left panels), 50 (middle panels), and 150 (right panels) messages per actor. The results for a model with a 1 and 2 dimensional latent variable are displayed in red and green, respectively.}
\label{vu_an3}
\end{figure}

\begin{figure}[h]
\centering
\makebox{\includegraphics[width=14cm]{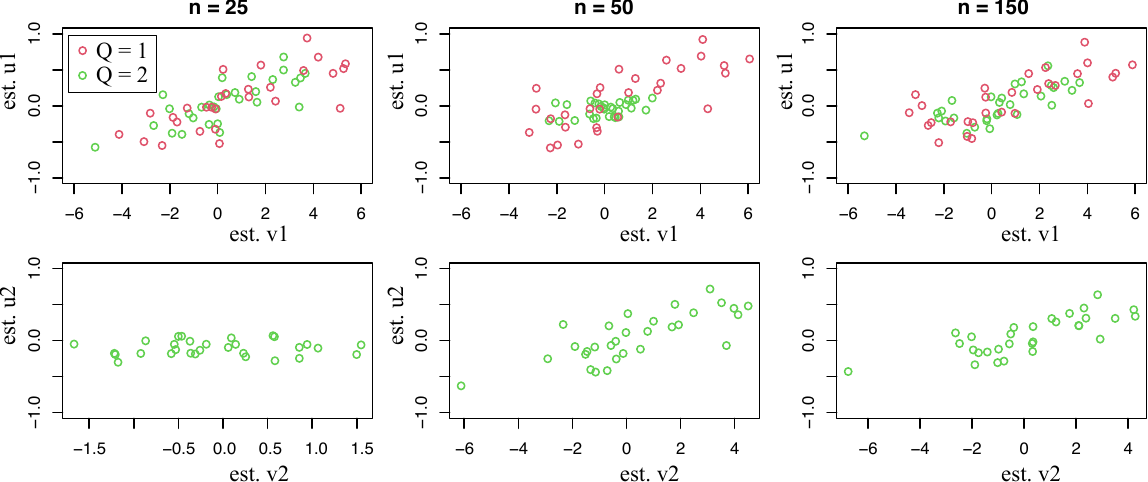}}
\caption{Population 2a (Table \ref{tabExtraSim}). Estimated latent variables of the actors as sender ($x$ axis) and as receiver ($y$ axis) for the first dimension (upper panels) and the second dimension (lower panels) in the case of 25 (left panels), 50 (middle panels), and 150 (right panels) messages per actor. The results for a model with a 1 and 2 dimensional latent variable are displayed in red and green, respectively.}
\label{vu_an4}
\end{figure}

\begin{figure}[h]
\centering
\makebox{\includegraphics[width=14cm]{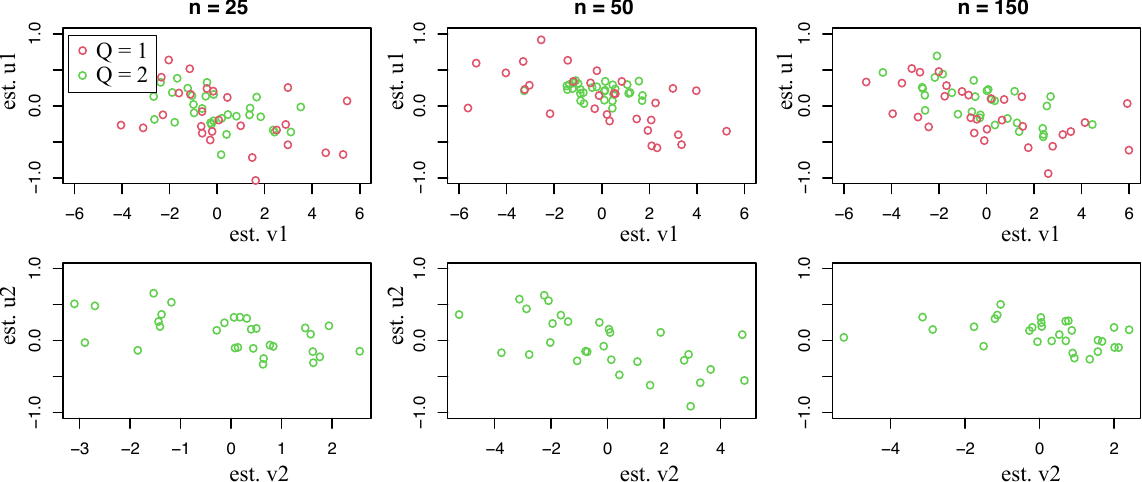}}
\caption{Population 2b (Table \ref{tabExtraSim}). Estimated latent variables of the actors as sender ($x$ axis) and as receiver ($y$ axis) for the first dimension (upper panels) and the second dimension (lower panels) in the case of 25 (left panels), 50 (middle panels), and 150 (right panels) messages per actor. The results for a model with a 1 and 2 dimensional latent variable are displayed in red and green, respectively.}
\label{vu_an5}
\end{figure}

\begin{figure}[h]
\centering
\makebox{\includegraphics[width=14cm]{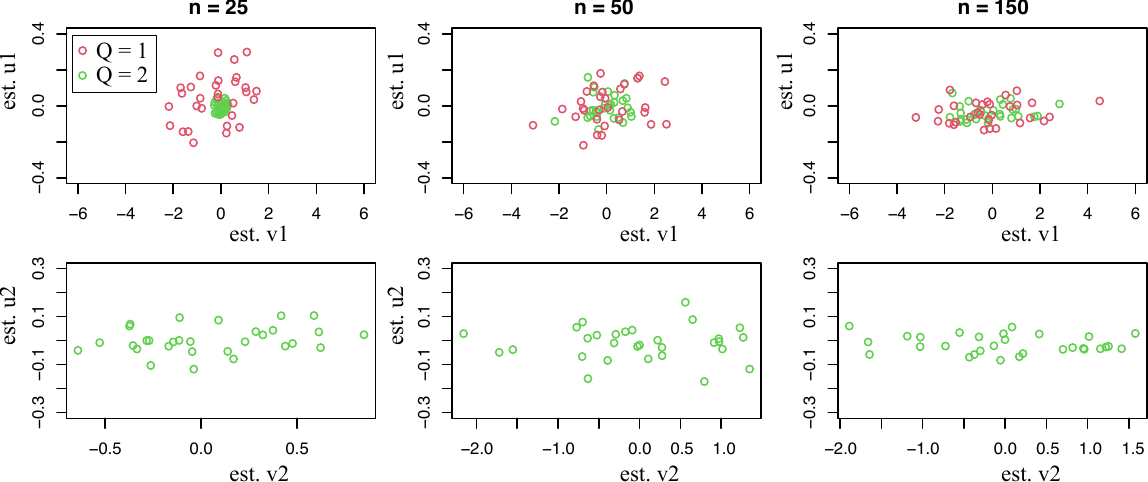}}
\caption{Population 3 (Table \ref{tabExtraSim}). Estimated latent variables of the actors as sender ($x$ axis) and as receiver ($y$ axis) for the first dimension (upper panels) and the second dimension (lower panels) in the case of 25 (left panels), 50 (middle panels), and 150 (right panels) messages per actor. The results for a model with a 1 and 2 dimensional latent variable are displayed in red and green, respectively.}
\label{vu_an6}
\end{figure}

\begin{figure}[h]
\centering
\makebox{\includegraphics[width=14cm]{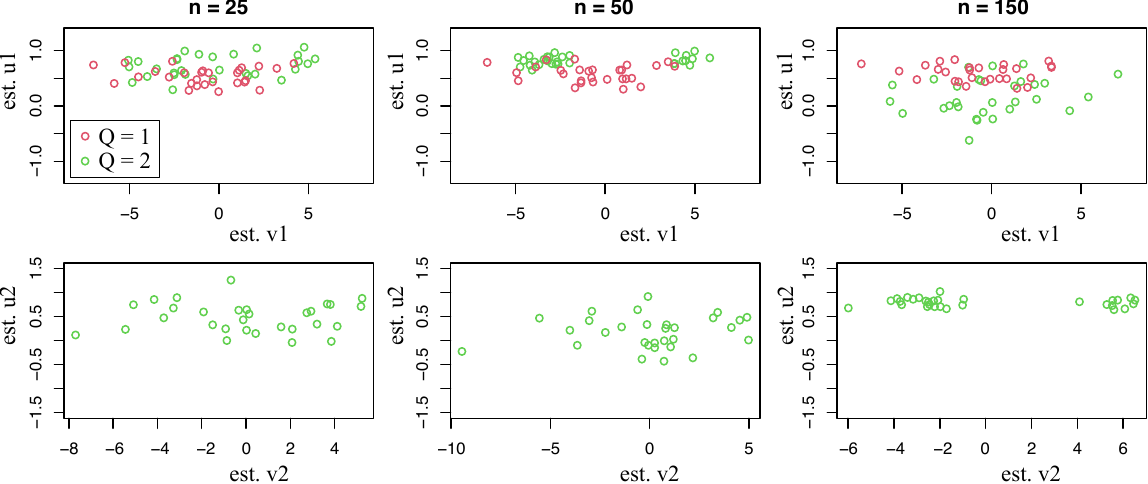}}
\caption{Population 4 (Table \ref{tabExtraSim}). Estimated latent variables of the actors as sender ($x$ axis) and as receiver ($y$ axis) for the first dimension (upper panels) and the second dimension (lower panels) in the case of 25 (left panels), 50 (middle panels), and 150 (right panels) messages per actor. The results for a model with a 1 and 2 dimensional latent variable are displayed in red and green, respectively.}
\label{vu_an7}
\end{figure}


\begin{figure}[h]
\centering
\makebox{\includegraphics[width=14cm]{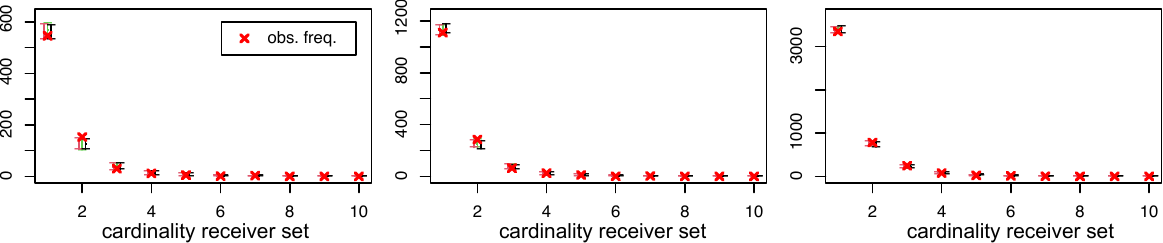}}
\caption{Population 0 (Table \ref{tabExtraSim}). Replicated frequencies of receiver sets with increasing cardinality in the case of 25 (left panels), 50 (middle panels), and 150 (right panels) messages per actor. The results for a model with 0, 1 and a 2 dimensional latent variable are displayed in black, red, and green, respectively. The observed frequencies are displayed as thick crosses.}
\label{cardi_an1}
\end{figure}

\begin{figure}[h]
\centering
\makebox{\includegraphics[width=14cm]{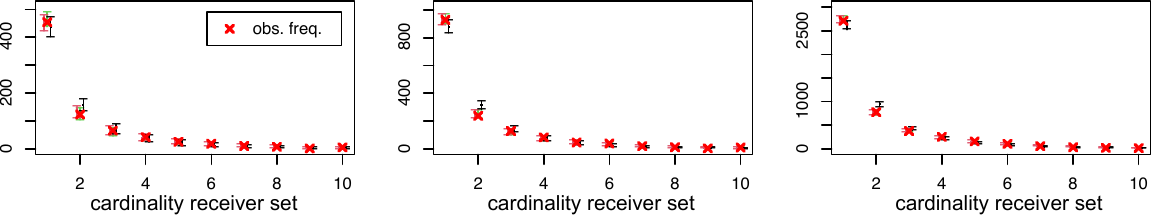}}
\caption{Population 1a (Table \ref{tabExtraSim}). Replicated frequencies of receiver sets with increasing cardinality in the case of 25 (left panels), 50 (middle panels), and 150 (right panels) messages per actor. The results for a model with 0, 1 and a 2 dimensional latent variable are displayed in black, red, and green, respectively. The observed frequencies are displayed as thick crosses.}
\label{cardi_an2}
\end{figure}

\begin{figure}[h]
\centering
\makebox{\includegraphics[width=14cm]{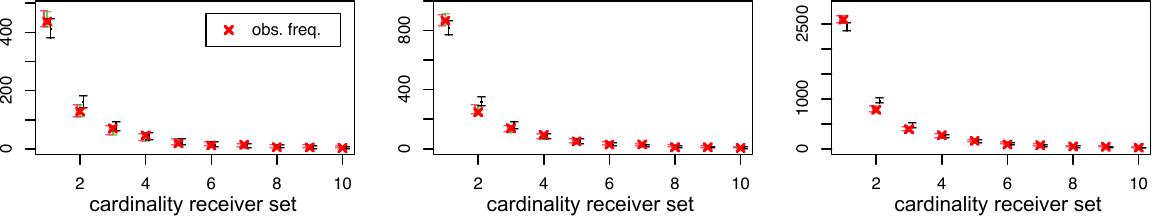}}
\caption{Population 1b (Table \ref{tabExtraSim}). Replicated frequencies of receiver sets with increasing cardinality in the case of 25 (left panels), 50 (middle panels), and 150 (right panels) messages per actor. The results for a model with 0, 1 and a 2 dimensional latent variable are displayed in black, red, and green, respectively. The observed frequencies are displayed as thick crosses.}
\label{cardi_an3}
\end{figure}

\begin{figure}[h]
\centering
\makebox{\includegraphics[width=14cm]{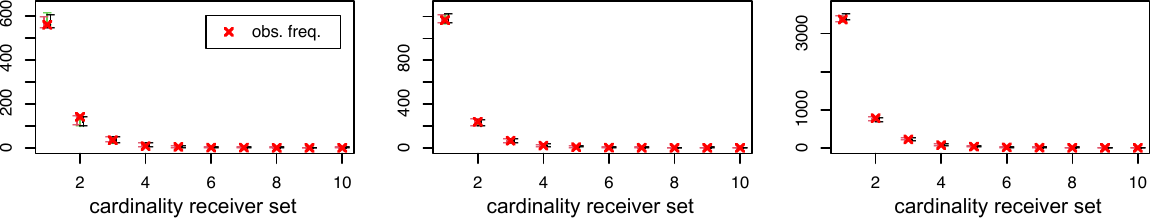}}
\caption{Population 2a (Table \ref{tabExtraSim}). Replicated frequencies of receiver sets with increasing cardinality in the case of 25 (left panels), 50 (middle panels), and 150 (right panels) messages per actor. The results for a model with 0, 1 and a 2 dimensional latent variable are displayed in black, red, and green, respectively. The observed frequencies are displayed as thick crosses.}
\label{cardi_an4}
\end{figure}

\begin{figure}[h]
\centering
\makebox{\includegraphics[width=14cm]{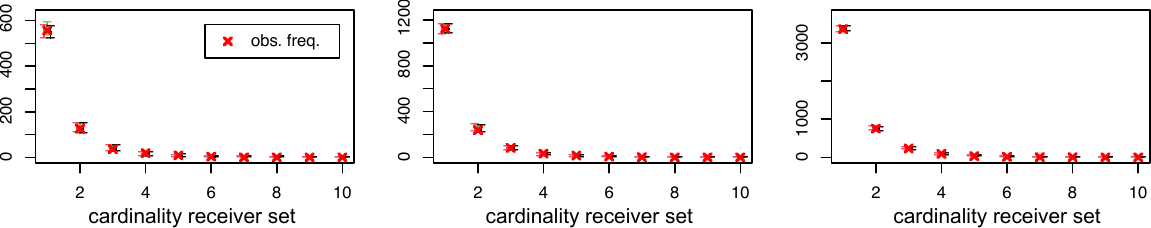}}
\caption{Population 2b (Table \ref{tabExtraSim}). Replicated frequencies of receiver sets with increasing cardinality in the case of 25 (left panels), 50 (middle panels), and 150 (right panels) messages per actor. The results for a model with 0, 1 and a 2 dimensional latent variable are displayed in black, red, and green, respectively. The observed frequencies are displayed as thick crosses.}
\label{cardi_an5}
\end{figure}

\begin{figure}[h]
\centering
\makebox{\includegraphics[width=14cm]{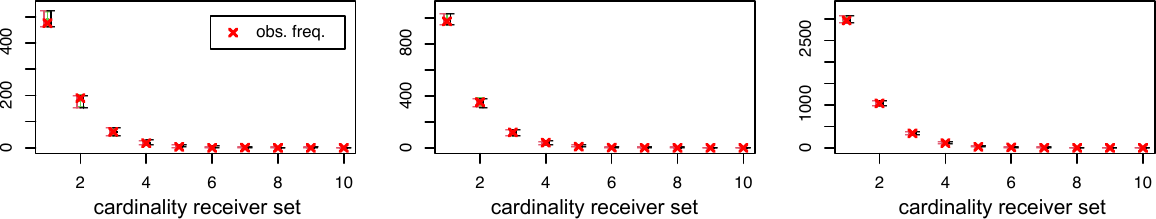}}
\caption{Population 3 (Table \ref{tabExtraSim}). Replicated frequencies of receiver sets with increasing cardinality in the case of 25 (left panels), 50 (middle panels), and 150 (right panels) messages per actor. The results for a model with 0, 1 and a 2 dimensional latent variable are displayed in black, red, and green, respectively. The observed frequencies are displayed as thick crosses.}
\label{cardi_an6}
\end{figure}

\begin{figure}[h]
\centering
\makebox{\includegraphics[width=14cm]{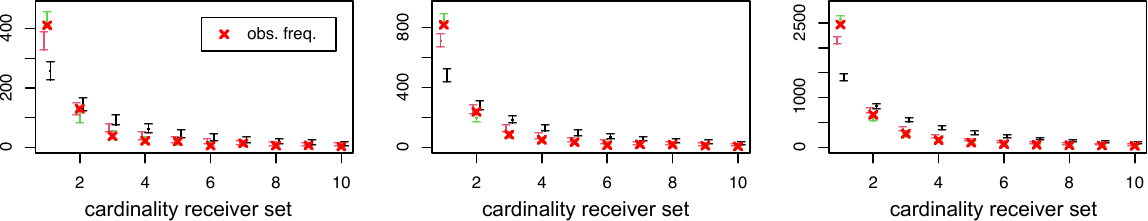}}
\caption{Population 4 (Table \ref{tabExtraSim}). Replicated frequencies of receiver sets with increasing cardinality in the case of 25 (left panels), 50 (middle panels), and 150 (right panels) messages per actor. The results for a model with 0, 1 and a 2 dimensional latent variable are displayed in black, red, and green, respectively. The observed frequencies are displayed as thick crosses.}
\label{cardi_an7}
\end{figure}

\begin{figure}[h]
\centering
\makebox{\includegraphics[width=14cm]{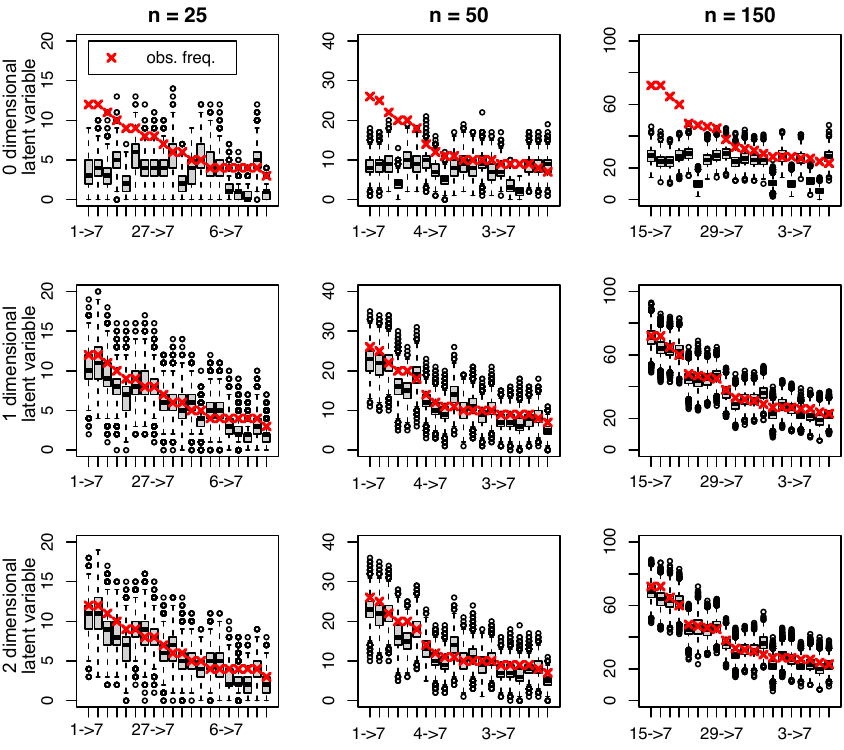}}
\caption{Population 0 (Table \ref{tabExtraSim}). Replicated frequencies of the most commonly observed pairs of senders and receiver sets in the case of 25 (left panels), 50 (middle panels), and 150 (right panels) messages per actor. The results for a model with 0, 1 and a 2 dimensional latent variable are displayed in the upper, middle, and lower panels, respectively. The observed frequencies are displayed as thick crosses.}
\label{recset_an1}
\end{figure}

\begin{figure}[h]
\centering
\makebox{\includegraphics[width=14cm]{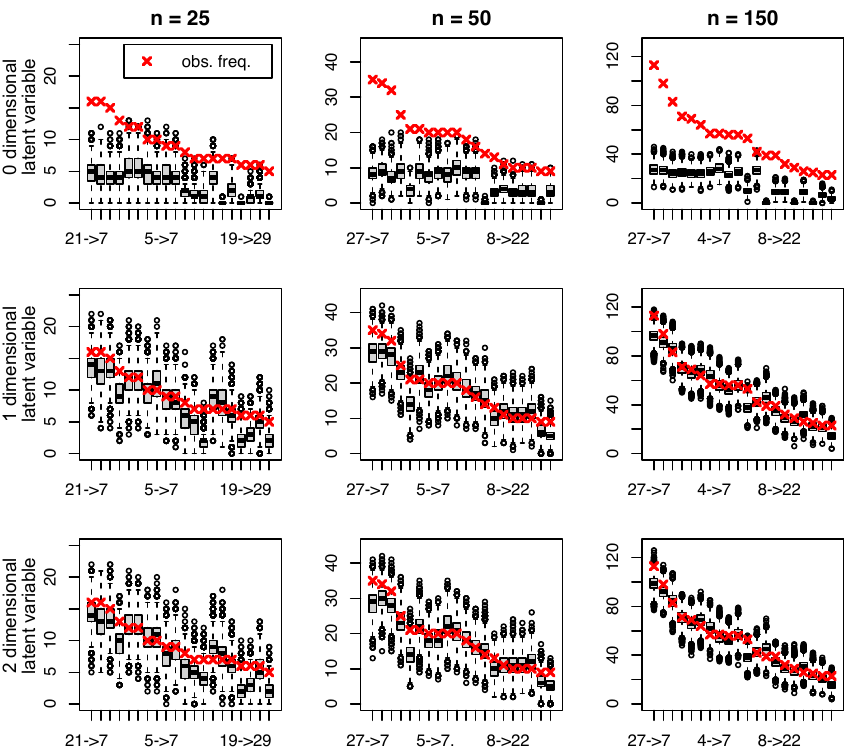}}
\caption{Population 1a (Table \ref{tabExtraSim}). Replicated frequencies of the most commonly observed pairs of senders and receiver sets in the case of 25 (left panels), 50 (middle panels), and 150 (right panels) messages per actor. The results for a model with 0, 1 and a 2 dimensional latent variable are displayed in the upper, middle, and lower panels, respectively. The observed frequencies are displayed as thick crosses.}
\label{recset_an2}
\end{figure}

\begin{figure}[h]
\centering
\makebox{\includegraphics[width=14cm]{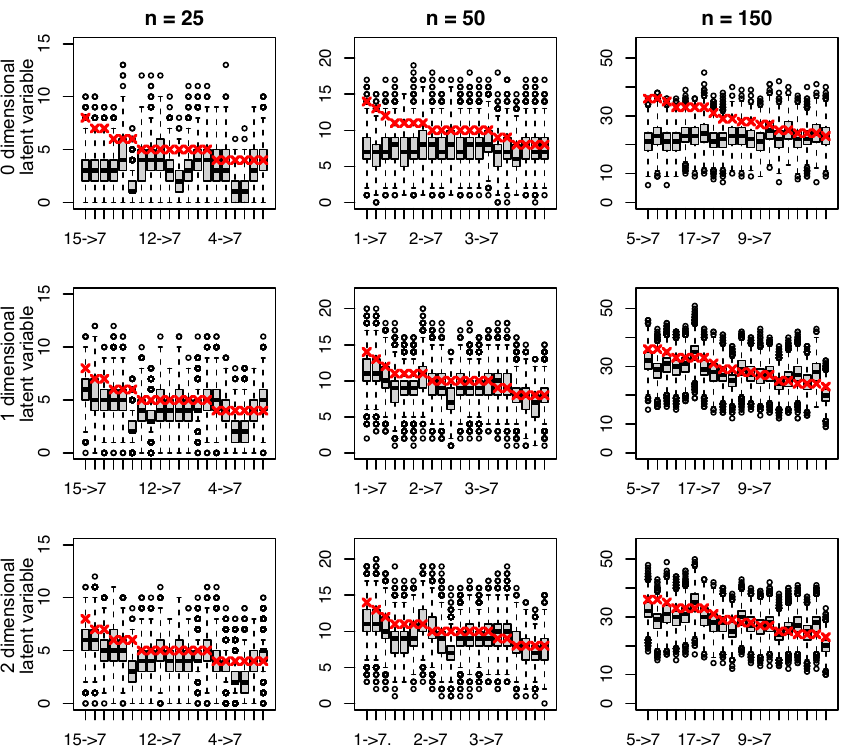}}
\caption{Population 1b (Table \ref{tabExtraSim}). Replicated frequencies of the most commonly observed pairs of senders and receiver sets in the case of 25 (left panels), 50 (middle panels), and 150 (right panels) messages per actor. The results for a model with 0, 1 and a 2 dimensional latent variable are displayed in the upper, middle, and lower panels, respectively. The observed frequencies are displayed as thick crosses.}
\label{recset_an3}
\end{figure}

\begin{figure}[h]
\centering
\makebox{\includegraphics[width=14cm]{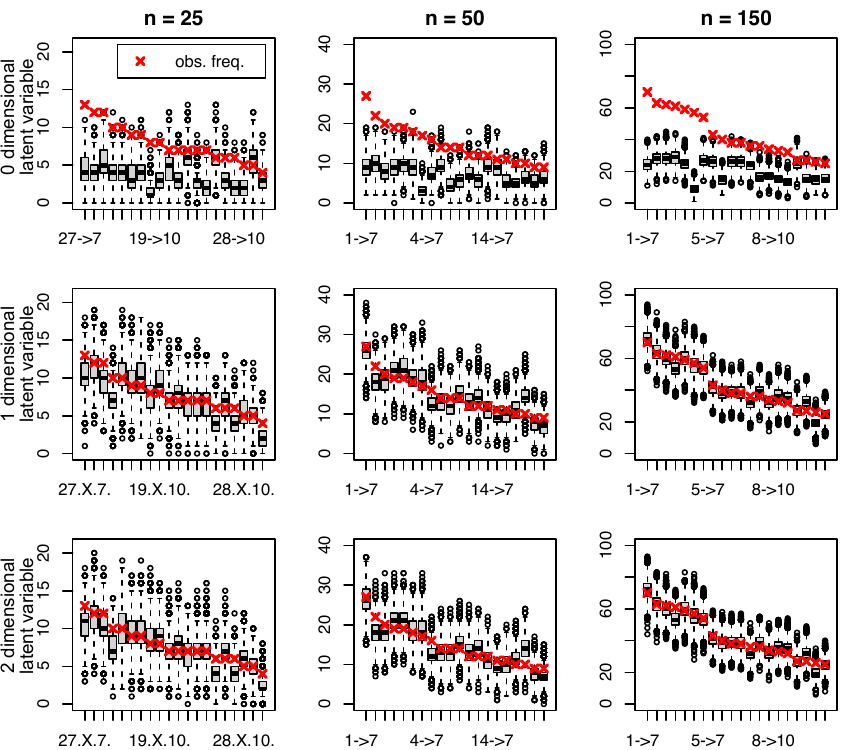}}
\caption{Population 2a (Table \ref{tabExtraSim}). Replicated frequencies of the most commonly observed pairs of senders and receiver sets in the case of 25 (left panels), 50 (middle panels), and 150 (right panels) messages per actor. The results for a model with 0, 1 and a 2 dimensional latent variable are displayed in the upper, middle, and lower panels, respectively. The observed frequencies are displayed as thick crosses.}
\label{recset_an4}
\end{figure}

\begin{figure}[h]
\centering
\makebox{\includegraphics[width=14cm]{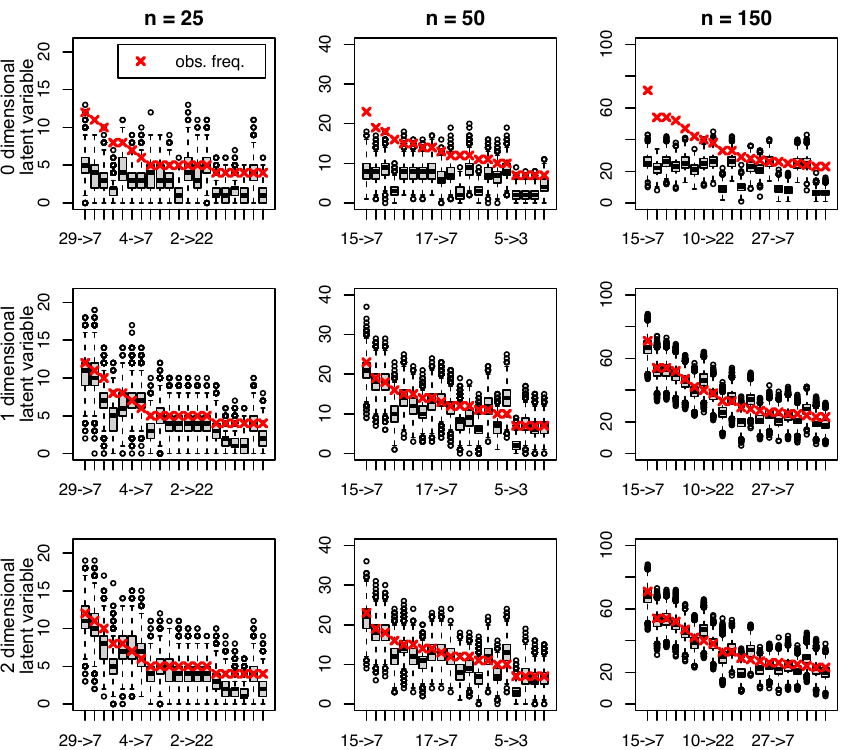}}
\caption{Population 2b (Table \ref{tabExtraSim}). Replicated frequencies of the most commonly observed pairs of senders and receiver sets in the case of 25 (left panels), 50 (middle panels), and 150 (right panels) messages per actor. The results for a model with 0, 1 and a 2 dimensional latent variable are displayed in the upper, middle, and lower panels, respectively. The observed frequencies are displayed as thick crosses.}
\label{recset_an5}
\end{figure}

\begin{figure}[h]
\centering
\makebox{\includegraphics[width=14cm]{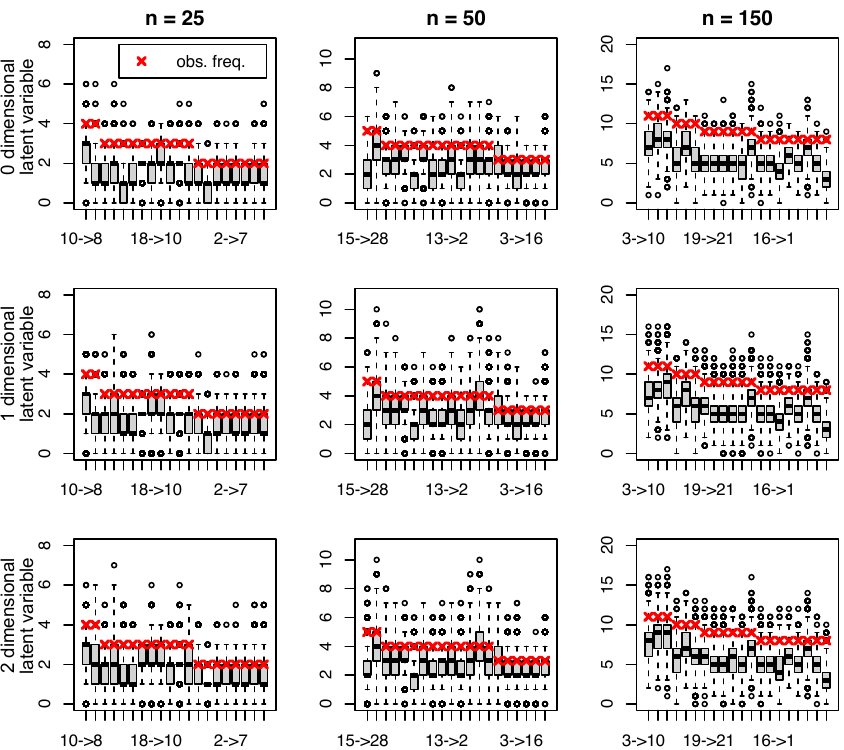}}
\caption{Population 3 (Table \ref{tabExtraSim}). Replicated frequencies of the most commonly observed pairs of senders and receiver sets in the case of 25 (left panels), 50 (middle panels), and 150 (right panels) messages per actor. The results for a model with 0, 1 and a 2 dimensional latent variable are displayed in the upper, middle, and lower panels, respectively. The observed frequencies are displayed as thick crosses.}
\label{recset_an6}
\end{figure}

\begin{figure}[h]
\centering
\makebox{\includegraphics[width=14cm]{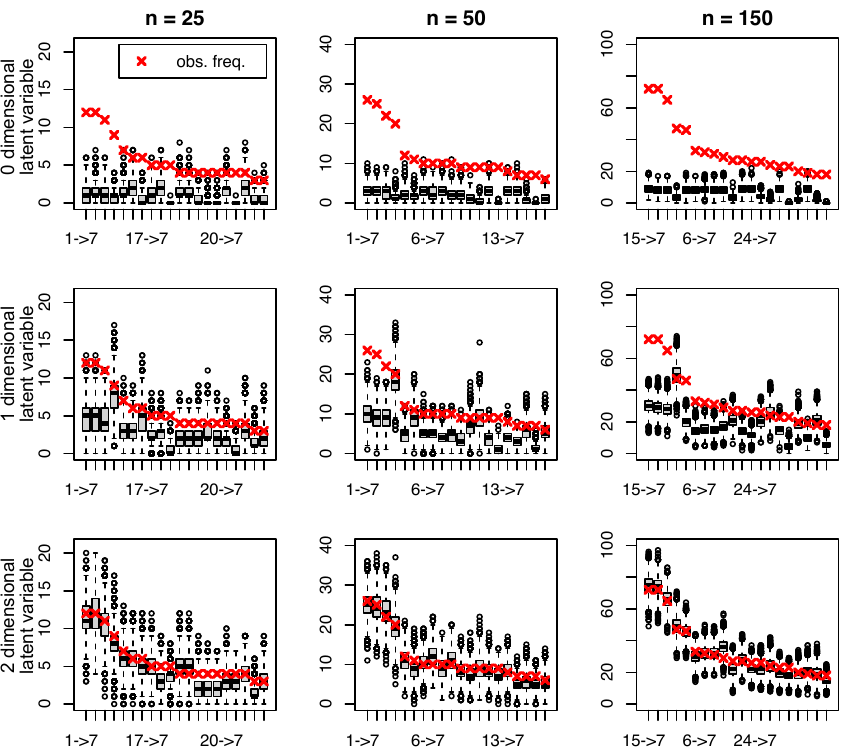}}
\caption{Population 4 (Table \ref{tabExtraSim}). Replicated frequencies of the most commonly observed pairs of senders and receiver sets in the case of 25 (left panels), 50 (middle panels), and 150 (right panels) messages per actor. The results for a model with 0, 1 and a 2 dimensional latent variable are displayed in the upper, middle, and lower panels, respectively. The observed frequencies are displayed as thick crosses.}
\label{recset_an7}
\end{figure}

\end{document}